\documentclass[longauth]{aa}  
\makeatletter
\renewcommand*\aa@pageof{, page \thepage{} of \pageref*{LastPage}}
\makeatother
\usepackage{adjustbox}
\usepackage{graphicx}
\usepackage{caption}
\usepackage{txfonts}
\usepackage{hyperref}
\usepackage{color}
\usepackage{float}

\begin{document}

    \titlerunning{The bright long-lived Type~II SN~2021irp powered by aspherical CSM interaction (I)}
    \authorrunning{T. M. Reynolds et al.}

   \title{The bright long-lived Type~II SN~2021irp powered by aspherical circumstellar material interaction (I): Revealing the energy source with photometry and spectroscopy.
   }

   \author{T. M. Reynolds\inst{1,2,3}
          \and
          T. Nagao\inst{1,4,5}
         \and
         R. Gottumukkala\inst{2,3}
         \and
         C. P. Guti\'errez\inst{6,7}
         \and
         T. Kangas\inst{8,1}
         \and
         T. Kravtsov\inst{1}
         \and
         H. Kuncarayakti\inst{1}
         \and
         K. Maeda\inst{9}
         \and
         N. Elias-Rosa\inst{10,7}
         \and
         M. Fraser\inst{11}
         \and
         R. Kotak\inst{1}
         \and
         S. Mattila\inst{1,12}
         \and
         A. Pastorello\inst{10}
         \and
         P. J. Pessi\inst{13}
         \and
         Y.-Z. Cai\inst{14,15,16}         
         \and
         J. P. U. Fynbo\inst{2,3}
         \and
         M. Kawabata\inst{17}
         \and
         P. Lundqvist\inst{13}
         \and
         K. Matilainen\inst{1}
         \and
         S. Moran\inst{1}
         \and
        A. Reguitti\inst{18,10}
         \and
         K. Taguchi\inst{17}
         \and         
         M. Yamanaka\inst{19}
          }

   \institute{
             Tuorla Observatory, Department of Physics and Astronomy, University of Turku, FI-20014 Turku, Finland
             \and
             Cosmic Dawn Center (DAWN)
             \and
             Niels Bohr Institute, University of Copenhagen, Jagtvej 128, 2200 København N, Denmark
             \and
             Aalto University Mets\"ahovi Radio Observatory, Mets\"ahovintie 114, 02540 Kylm\"al\"a, Finland
             \and
             Aalto University Department of Electronics and Nanoengineering, P.O. BOX 15500, FI-00076 AALTO, Finland
             \and
             Institut d'Estudis Espacials de Catalunya (IEEC), Edifici RDIT, Campus UPC, 08860 Castelldefels (Barcelona), Spain
             \and
             Institute of Space Sciences (ICE, CSIC), Campus UAB, Carrer de Can Magrans, s/n, E-08193 Barcelona, Spain
             \and
             Finnish Centre for Astronomy with ESO (FINCA), FI-20014 University of Turku, Finland
             \and
             Department of Astronomy, Kyoto University, Kitashirakawa-Oiwake-cho, Sakyo-ku, Kyoto 606-8502, Japan
             \and
             INAF – Osservatorio Astronomico di Padova, Vicolo dell'Osservatorio 5, I-35122 Padova, Italy
             \and
             UCD School of Physics, L.M.I. Main Building, Beech Hill Road, Dublin 4,D04 P7W1, Ireland
             \and
             School of Sciences, European University Cyprus, Diogenes Street, Engomi, 1516, Nicosia, Cyprus
             \and 
             The Oskar Klein Centre, Department of Astronomy, Stockholm University, Albanova University Center, SE 106 91 Stockholm, Sweden
             \and
             Yunnan Observatories, Chinese Academy of Sciences, Kunming 650216, P.R. China
             \and
             International Centre of Supernovae, Yunnan Key Laboratory, Kunming 650216, P.R. China
             \and 
             Key Laboratory for the Structure and Evolution of Celestial Objects, Chinese Academy of Sciences, Kunming 650216, P.R. China
             \and
             Okayama Observatory, Kyoto University, 3037-5 Honjo, Kamogatacho, Asakuchi, Okayama 719-0232, Japan
             \and
             INAF – Osservatorio Astronomico di Brera, Via E. Bianchi 46, I-23807 Merate (LC), Italy
             \and
             Amanogawa Galaxy Astronomy Research Center (AGARC), Graduate School of Science and Engineering, Kagoshima University, 1-21-35 Korimoto, Kagoshima, Kagoshima 890-0065, Japan
             }

   \date{...}

 
  \abstract
   {Some core-collapse supernovae (CCSNe) are too luminous and radiate too much total energy to be powered by the release of thermal energy from the ejecta and radioactive-decay energy from the synthesised $^{56}$Ni/$^{56}$Co. A source of additional power is the interaction between the supernova (SN) ejecta and a massive circumstellar material (CSM). This is an important power source in Type IIn SNe, which show narrow spectral lines arising from the unshocked CSM, but not all interacting SNe show such narrow lines.}
   {We present photometric and spectroscopic observations of the hydrogen-rich SN~2021irp, which is both luminous, with $M_{o} < -19.4$ mag, and long-lived, remaining brighter than $M_{o} = -18$ mag for $\sim$ 250~d. We show that an additional energy source is required to power such a SN, and determine the nature of the source. We also investigate the properties of the pre-existing and newly formed dust associated with the SN.}
   {Photometric observations 
   show that the luminosity of the SN is an order of magnitude higher than typical Type II SNe and persists for much longer. We detect a infrared excess attributed to dust emission. Spectra show multi-component line profiles, an \ion{Fe}{II} pseudo-continuum and a lack of absorption lines, all typical features of Type IIn SNe. We detect a narrow (< 85 kms$^{-1}$) P-Cygni profile associated with the unshocked CSM. An asymmetry in emission line profiles indicates dust formation occurring from 250-300~d. Analysis of the SN blackbody radius evolution indicates asymmetry in the shape of the emitting region.}
   {We identify the main power source of SN~2021irp as extensive interaction with a massive CSM, and that this CSM is distributed asymmetrically around the progenitor star. The infrared excess is explained with emission from newly formed dust although there is also some evidence of an IR echo from pre-existing dust at early times.}
   {}


   \maketitle
%

\section{Introduction}

Hydrogen-rich (Type~II) core-collapse supernovae (SNe) are explosions of massive stars, characterised by bright transients that show hydrogen Balmer lines in their spectra. The most common subclass of SNe in this category is Type IIP. Type~IIP SNe are explosions of red supergiant stars whose Zero-Age-Main-Sequence masses are between $\sim8$ and $\sim18~M_{\odot}$ \citep[see e.g.][]{VanDyk2003,Smartt2004,Smartt2015a}; although see e.g. \citet{Beasor2024} who argue for somewhat higher masses. After an initial rise to peak and a short decline period, they then show relatively constant brightness ($-15 > M_V > -18$ mag) for the first $\sim 100$ days \citep[the plateau phase; e.g.][]{Barbon1979,Anderson2014}, which is followed by an exponential decline (the tail phase) after a short period of steep decline \citep[][]{Zampieri2017}. The radiation during the plateau phase originates mainly from the thermal energy deposited by the explosion shock, with an additional contribution by the decay of $^{56}$Ni/$^{56}$Co, while the power in the tail phase is dominated by the decay of $^{56}$Co. The typical total energy radiated in the plateau phase is $\sim 10^{49}$ erg, inferred from the typical bolometric luminosity of $\sim 10^{42}$ erg~s$^{-1}$ and the typical plateau length of $\sim 10^{7}$ s \citep[e.g.,][]{Valenti2016,Martinez2022}.

There are some extremely luminous Type~II SNe whose total radiated energy is much larger than the typical values for Type IIP SNe. Such bright Type II SNe need some other energy source in addition to the thermal energy and the radioactive decay of $^{56}$Ni / $^{56}$Co. Many of these luminous SNe are so-called Type~IIn SNe \citep[][for reviews]{Smith2017a,Fraser2020}. The characteristic observational feature of Type~IIn SNe is narrow Balmer lines in their spectra \citep{Schlegel1990}, which are believed to originate in the circumstellar material (CSM) ionised/excited by high-energy photons from the CSM interaction shocks. 
Although these SNe are typically luminous, there is a large diversity in their brightness, with the typical $r$-band peak absolute magnitude being around $-19.18 \pm 1.32$ mag \citep[][]{Nyholm2020}. These large luminosities are interpreted as the results of the extensive CSM interaction which converts some of the kinetic energy of the SN ejecta to optical radiation. The total mass of CSM required can be larege, with the well-studied luminous and long-lived Type IIn SN 2010jl, which is mainly powered by strong CSM interaction, requiring several solar masses of CSM \citep[e.g.,][]{Fransson2014}.

The majority of Type~II superluminous SNe (SLSNe), which reach $\lesssim -20$ mag \citep[see e.g.][]{GalYam2012,GalYam2019,Kangas2022,Moriya2024,Pessi2024}, show prominent Balmer emission lines with a narrow core on a broader base and are thus understood as extreme cases of classical Type~IIn SNe, i.e., SNe with even stronger CSM interaction. On the other hand, some Type~II SLSNe show only broad Balmer lines and lack narrow ones \citep{Gezari2009,Miller2009,Inserra2018,Kangas2022}. These bright SNe are also suggested to be interaction-powered SNe with different properties of CSM \citep[][]{Moriya2012,McDowell2018}, although some other energy sources, e.g., a magnetar central engine, a fallback accretion central engine, or a large amount of $^{56}$Ni, have been proposed \citep[e.g.][]{Inserra2018,Terreran2017}. In some rare Type II SLSNe, the total radiated energy is larger than the total kinetic energy for a typical core-collapse SN, implying that CSM interaction alone is not a sufficient power source \citep{Kangas2022,Pessi2024}.

There are other Type II SNe that are more luminous and energetic than typical Type IIP SNe, are not SLSNe, and do not show the narrow lines that unambiguously identify CSM interaction. These include the rare class of luminous Type IIL SNe, which have absolute magnitudes between -18 and -20 mag, and exhibit a linear decline \citep[e.g][]{Kangas2016,Reynolds2020,Pessi2023}. For many of these SNe, CSM interaction is invoked to explain their large peak luminosity and radiated energy as well as the long-lasting relatively blue and featureless spectrum seen at early times. Direct evidence of CSM interaction can be seen in the radio \citep{Lundqvist1988,Weiler1991}, X-ray \citep{Immler2005} and infrared (IR) emission \citep{Dwek1983} from the well-studied luminous and linear SN~1979C. The fast decline for these SNe implies that any CSM interaction is less intense than that in long-lasting Type IIn SNe, such as SN~2010jl \citep{Fransson2014} and SN 2015da \citep{Tartaglia2020}. Similarly to Type II SLSNe, other energy sources have been invoked for some objects \citep[see e.g.][]{Bose2018}.

We present the bright, long-lived Type~II SN~2021irp, which shows a broad Balmer line dominated spectrum, and discuss its energy source. The paper is structured as follows. In Sect.~\ref{sec:Obs_data_red}, we describe the discovery of the SN, its observational parameters, and the acquisition and reduction of our data. In Sect.~\ref{sec:host_galaxy}, we discuss some properties of the host galaxy and derive a host extinction value. In Sect.~\ref{sec:PhotAnalysis} and Sect.~\ref{sec:spectra}, we describe and analyse the photometric and spectroscopic properties of the SN, respectively. Section \ref{sec:discussion} discusses the required energy source for SN~2021irp and the presence and effect of pre-existing and newly formed dust. Section~\ref{sec:conclusions} summarises our work and presents our conclusions.


\section{Observations and data reduction}
\label{sec:Obs_data_red}
\subsection{Discovery, classification and observational parameters.}

SN 2021irp (a.k.a ATLAS21lhv and Gaia21ekq) was discovered by the Asteroid Terrestrial-impact Last Alert System (ATLAS; \citet{Tonry2018,SmithK2020}) on 9 April 2021 (MJD = 59313.2) at a magnitude of m$_c$ = 18.02 and subsequently reported to the Transient Name Server (TNS\footnote{\url{https://www.wis-tns.org/}}). The authenticity of the detection was confirmed by further ATLAS detections, but the object then passed behind the sun, preventing further observations or classification. The transient was next detected by ATLAS on 27 July 2021 (MJD = 59421.6; 108 days later) but remained unclassified until 8 October 2021 (MJD = 59495.3), when it was spectroscopically classified as a Type II SN \citep{Classification}.

\begin{figure}
   \centering
   \includegraphics[width=0.5\textwidth]{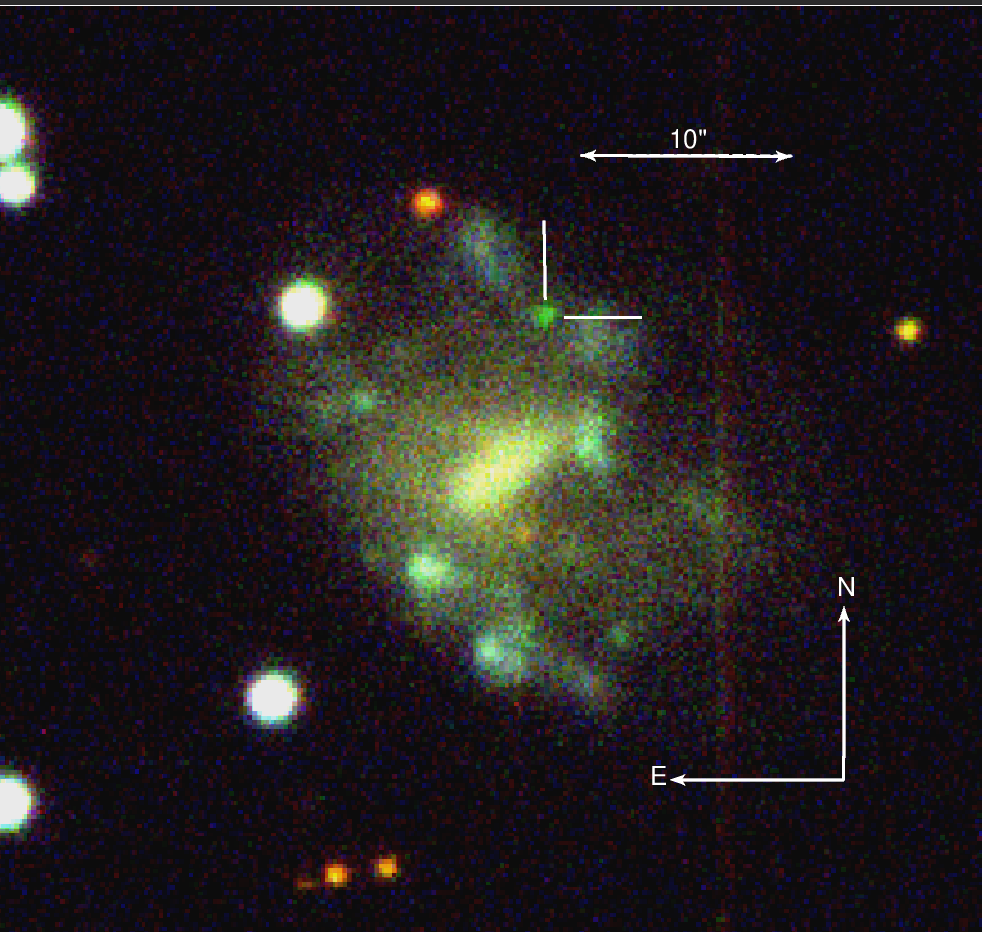}
      \caption{Combined $Bri$ images of the field of SN~2021irp obtained with the NOT+ALFOSC. We used our best quality images: the $B$ and $i$ images are taken after the SN has entirely faded, while the $r$ image is a late time (499~d) deep image in which the SN is still visible and indicated by the tick marks. The faint line on the right side of the image is a saturation artefact from a bright star.}
\label{fig:SN_host_image}
\end{figure}

The last non-detection of the SN before detection by ATLAS was on 3 April 2021 (MJD = 59307.3) with a 3 sigma detection limit of m$_o \geq$ 19.84. Additionally, a ``null" detection of SN 2021irp was reported by the Gaia Alerts survey on 5 April 2021 (MJD = 59309.6). Due to the uncertain nature of this non-detection\footnote{See \url{http://gsaweb.ast.cam.ac.uk/alerts/docs/}}, here we adopt the midpoint of the ATLAS non-detection and first detection as the explosion epoch, namely 6 April 2021, MJD=59310.3 $\pm 3$. All phases given are in rest frame days with respect to this explosion epoch unless specified otherwise.

We take the foreground Milky Way (MW) extinction towards SN~2021irp to be $A_V$ = 1.193 mag \citep[from \citet{Schlafly2011}, via the NASA/IPAC Infrared Science Archive Galactic Dust Reddening and Extinction tool][]{IRSA_dust}. As discussed below in Sect. \ref{sect:host_extinction}, we infer an additional extinction of $A_V = 0.12 \pm 0.04$ mag arising within the host from measurements of resolved Na~{\sc i}~D absorption lines. We therefore take the total extinction towards SN~2021irp as $A_V = 1.313 \pm 0.04$ mag. We correct all optical and NIR photometry using the \citet{Cardelli1989} extinction law with this total extinction and assuming R$_{V} = 3.1$. We correct the MIR photometry using the \citep{Fitzpatrick1999} extinction law with the coefficients from \citet{Yuan2013}. Photometry in ATLAS $co$ and SDSS $riz$ bands is presented in the AB system, while the remaining photometry is presented in the Vega system. We adopt z = $0.0195\pm0.0001$ as the redshift from measurements of the position of the narrow host Na~{\sc i}~D absorption lines and the narrow H$\alpha$ emission line observed in our spectra. This corresponds to a distance of $85.3\pm0.4$ Mpc (distance modulus = $34.654\pm0.002$~mag) assuming a $\Lambda$CDM cosmology and taking H$_{0}$ = 69.6 km s$^{-1}$Mpc$^{-1}$ and $\Omega_{m}$ = 0.286 \citep{Bennett2014}. 

We show the SN and host galaxy in Fig. \ref{fig:SN_host_image}. SN~2021irp is located at the coordinates RA = 05:23:27.46, Dec = +17:04:40.10, and the host galaxy of SN~2021irp is a known Sloan Digital Sky Survey \citep[SDSS;][]{SDSSdr17} galaxy, with SDSS coordinates RA = 05:23:27.61, Dec=+17:04:32.51. The SN is 7.9" separated from the nucleus, corresponding to 3.1 kpc at our adopted redshift. The SN lies in a structure to the north and west of the nucleus that is potentially a spiral arm, which contains bluer regions that could be associated with star formation, although SN~2021irp does not appear to lie within such a region. We discuss the host in more detail in Sect. \ref{sec:host_galaxy}.

\subsection{Photometry}

\subsubsection{Optical}

Optical imaging of SN~2021irp was obtained with the 2.56m Nordic Optical Telescope (NOT\footnote{\url{http://www.not.iac.es}}) using the Alhambra Faint Object Spectrograph and Camera (ALFOSC) instrument as part of the NOT Un-biased Transient Survey 2 (NUTS2\footnote{\url{https://nuts.sn.ie/}}) program, primarily with the Sloan $ri$ and the Johnson-Cousins $BV$ filters. Photometry was additionally obtained as a part of the routine operations of the ATLAS survey in the cyan ($c$) and orange ($o$) bands, which approximately correspond to $g+r$ and $r+i$ respectively; and the Gaia Alerts project, which uses the $Gaia~(G)$ filter corresponding approximately to $g+r+i$. 

The NOT+ALFOSC images were reduced with the {\sc alfoscgui}\footnote{FOSCGUI is a graphical user interface aimed at extracting SN spectroscopy and photometry obtained with FOSC-like instruments. It was developed by E. Cappellaro. A package description can be found at \url{sngroup.oapd.inaf.it/foscgui.html}} pipeline. The reduction steps consisted of bias and overscan subtraction, and flat-fielding. Photometry was performed using the {\sc autophot} pipeline \citep{Brennan2022}. The pipeline removes artefacts produced by cosmic rays; corrects the image world coordinate system using an implementation of {\sc astrometry.net} \citep{Lang2010}; builds a PSF model from sources in the image; calibrates the image with a suitable catalogue by measuring the magnitude of field stars through PSF fitting; measures the magnitude of the transient also using PSF fitting; and finally calculates the uncertainty through source injection and recovery. A complete description of the software is given in \citet{Brennan2022}. The NOT+ALFOSC $BVgriz$ images were calibrated to the Pan-STARRS1 (PS1) catalogue \citep{Chambers2016,Flewelling2020}, where we used the equations provided in \citet{Tonry2012} to estimate magnitudes for field stars in the $B$ and $V$ band from the PS1 measurements. We obtained deep template imaging with the NOT+ALFOSC at +673~d after the explosion epoch, where the SN is not detected. Before measuring the magnitude of the SN, we subtracted the template images from those with the SN using the {\sc hotpants}\footnote{ {\sc hotpants} is available at \url{https://github.com/acbecker/hotpants}} algorithm, an implementation of the image subtraction algorithm by \cite{Alard1998}, after aligning the images using {\sc astroalign} \citep{Astroalign}.

ATLAS photometry was obtained through the ATLAS forced photometry server\footnote{\url{https://fallingstar-data.com/forcedphot/}}. We discard ATLAS detections with a signal-to-noise of less than three, and instead obtain 3-$\sigma$ upper limits. Gaia photometry was downloaded from the Gaia Science Alerts webpages\footnote{\url{http://gsaweb.ast.cam.ac.uk/alerts/alert/Gaia21ekq/}}. We adopt uncertainties for the Gaia photometry following the analysis in Sect. 3.5 of \citet{Hodgkin2021}. For both ATLAS and NOT+ALFOSC photometry, individual measurements were binned on nights with multiple observations, and the standard error of the mean of the binned measurements was adopted as the uncertainty. We additionally discard binned ATLAS detections with uncertainty > 0.2 mag, and a single $o$-band outlier detection, that is inconsistent with the ALFOSC data. A log of the photometric observations is given in Table \ref{tab:photometry_table}.

To check for pre-SN transient activity at the SN location we queried the ATLAS forced photometry server for all available observations before the SN explosion date. The data was sigma-clipped and stacked nightly using publically available software \citep{Young_plot_atlas_fp}. The earliest data available was obtained on 12th Sept 2015 (MJD = 57277.6), and the SN field was observed on 313 occasions between then and the SN explosion 1991~d later, with consistent observations every few days - few weeks except during the periods where the SN was not visible from the ATLAS site. There are no detections, defined as observations with signal to noise > 3, during this period. The median 3$\sigma$ upper limits across all the data in the two separate filters are $M_c = -15.37$~mag  and $M_o = -15.36$~mag, with the assumed distance and extinction for SN~2021irp. There is good coverage of the SN for the 244~d directly before the SN explosion, which is the most likely period to observe a precursor \citep{Strotjohann2021}, and these images are deeper, with the median limiting magnitude $M_c = -15.03$~mag. Precursor outbursts with magnitudes $\leq-13$ are not uncommon for Type IIn SNe \citep{Ofek2014}, with a recent study by \citet{Strotjohann2021} finding that $\sim$ 25\% of Type IIn SNe have such an outburst within the 90 days before explosion. Additionally, \citet{Strotjohann2021} find that $\sim$ 2\% of SNe with peak brightnesses $<-18.5$ such as SN~2021irp exhibit luminous outbursts $\leq-16$~mag. Our observations can rule out only the more luminous precursors and not those at the lower end of the observed luminosity distribution.


\subsubsection{Infrared}

Near-infrared (NIR) imaging of SN~2021irp was also obtained with the NOT, using the NOTCam instrument and the $JHKs$ filters. The NOTCam data were reduced using a slightly modified (e.g. to include the full field of view) version of the NOTCam {\sc quicklook}\footnote{\url{https://www.not.iac.es/instruments/notcam/quicklook.README}} v2.5 reduction package. The reduction process included flat-field correction, distortion correction, bad pixel masking, sky subtraction, and finally, stacking of the dithered images. Photometry was again performed with the {\sc autophot} pipeline, and the magnitudes were calibrated using the Two Micron All Sky Survey \citep[2MASS;][]{Skrutskie2006} catalogue.

Mid-infrared (MIR) imaging of the SN site was obtained by the Wide-field Infrared Survey Explorer (\textit{WISE}) satellite as part of the Near-Earth Object Wide-field Infrared Survey Explorer reactivation mission \citep[NEOWISE;][]{Mainzer2014}, which surveys the entire sky with the $W1$ and $W2$ filters (3.4~$\mu$m and 4.6~$\mu$m respectively) every 6 months. We obtained the NEOWISE photometric measurements of SN~2021irp from the NEOWISE-R Single Exposure (L1b) Source Table\footnote{\url{https://irsa.ipac.caltech.edu/cgi-bin/Gator/nph-scan?mission=irsa&submit=Select&projshort=WISE}} \citep{NEOWISE}, and adopt the median value of the individual measurements as the magnitude. For the uncertainty, we adopted the standard error of the mean of the individual measurements with additional 0.0026 mag and 0.0061 mag uncertainties added in quadrature to the $W1$ and $W2$ measurements, respectively, which represent the RMS residuals found in the photometric calibration of WISE during the survey period. 

\subsection{Spectroscopy}

We obtained 9 optical spectra of SN~2021irp with the NOT+ALFOSC; the FOcal Reducer/low-dispersion Spectrograph 2 (hereafter FORS2; \citealt{Appenzeller1998}) mounted at the Cassegrain focus of the Very Large Telescope (VLT) UT1 telescope at the Paranal observatory; and the Kyoto Okayama Optical Low-dispersion Spectrograph with optical-fiber Integral Field Unit \citep[KOOLS=IFT;][]{Matsubayashi2019} mounted on the 3.8-m Seimei telescope \citep[][]{Kurita2020} at the Okayama Observatory. A log of the spectroscopic observations is given in Table \ref{tab:Spectra_log}.

The ALFOSC spectra observed with Grism 4 were reduced with the {\sc alfoscgui} pipeline, which uses standard {\sc iraf}\footnote{https://iraf-community.github.io/} tasks to perform overscan, bias and flat-field corrections, as well as removing cosmic ray artefacts using {\sc lacosmic} \citep{vanDokkum2001}. One-dimensional spectra were extracted using the {\sc apall} task, and wavelength calibration was performed by comparison with arc lamps and corrected, if necessary, by comparison with skylines. The spectra were flux-calibrated against a sensitivity function derived from standard stars observed on the same night and absolutely calibrated using our photometric measurements. In order to estimate the flux of the SN at the epoch of spectroscopic observations, we interpolated the observed photometry with Gaussian processing, as described in Sect. \ref{sec:PhotAnalysis}. The single NOT+ALFOSC spectrum observed with Grism 7 was reduced using the {\sc pypeit}\footnote{\url{https://github.com/pypeit/PypeIt}} Python package \citep{pypeit2020}. This performed all the reduction steps similarly to as described above, with some additional features (e.g. calculating spectral tilts from the arc lamp exposures).
The VLT+FORS2 spectra were reduced using the ESOReflex \citep[][]{Freudling2013} pipeline following standard procedures.
The KOOLS=IFT spectra were reduced with the standard procedures that are developed for KOOLS=IFT data \footnote{\url{http://www.o.kwasan.kyoto-u.ac.jp/inst/p-kools/reduction-201806/index.html}}, using IRAF. The procedures include overscan, bias, flat-field corrections, and cosmic-ray removal. Extraction of a one-dimensional spectrum and sky subtraction were performed. Wavelength and flux calibrations were performed using arc lamps and a standard star. All the spectra, along with the photometric tables, are publicly available on the Weizmann Interactive Supernova data REPository\footnote{\url{ https://www.wiserep.org/object/19464}} \citep[WISeREP][]{Yaron2012}.


\section{Host galaxy}
\label{sec:host_galaxy}

The host galaxy of SN~2021irp is WISEA~J052327.68+170431.2 / SDSS~J052327.62+170432.4. We measured magnitudes for the host using the {\sc hostphot}\footnote{\url{https://github.com/temuller/hostphot}} package \citep{MullerBravo2022b}. Using utilities available as part of the package, we collected the imaging available from the PS1, SDSS, and 2MASS surveys; identified foreground stars, cross-matched them with the Gaia catalogue and masked them; and then performed aperture photometry in the images with a 6~kpc / 15.2" radius aperture centred on the galaxy nucleus. We additionally took the aperture magnitudes in a 16.5" radius aperture from the AllWISE catalogue. The photometry is presented in Table \ref{tab:host_mags}. The host luminosity is typical of a Type IIn SN host, with an absolute $g$-band magnitude of $\sim -19$, which is close to the average for the sample of Type IIn host galaxies presented in \citet{Nyholm2020}. As shown in Fig. \ref{fig:SN_host_image}, the structure of the galaxy appears disturbed, lacking clear spiral arms, although there is a central bar feature and blue knots likely indicating star-forming regions. Note that the significant MW extinction causes the host to appear redder than its intrinsic colour.

\begin{table}
\caption{Parameters derived from our SED fitting with {\sc bagpipes}.}
\centering
\begin{adjustbox}{width=0.5\textwidth}
\begin{tabular}{c c c c c} \hline\hline 
Stellar mass & SFR & sSFR & A$_{V}$ & Z \\ 
log($M_{*} / M_{\odot}$) & $M_{\odot}$~yr$^{-1}$ & log(sSFR / yr$^{-1}$  ) & mag & $Z/Z_{\odot}$ \\ 
\hline
$9.19^{+0.04}_{-0.04}$ & $0.51^{+0.03}_{-0.03}$ & $-9.48^{+0.04}_{-0.05}$ & $0.19^{+0.02}_{-0.01}$ & $0.34^{+0.05}_{-0.05}$\\
\end{tabular}
\end{adjustbox}
\label{tab:host_fitting_params}
\end{table}

To estimate physical parameters for the host galaxy, we modelled the photometric spectral energy distribution (SED) using the SED fitting package Bayesian Analysis of Galaxies for Physical Inference and Parameter EStimation \citep[{\sc bagpipes},][]{Carnall2018}. {\sc bagpipes} performs Bayesian fitting to photometric and spectroscopic data with model spectra of galaxies generated using stellar population synthesis models \citep{Bruzual2003} with user-specified dust attenuation laws, an intergalactic medium attenuation model from \cite{Inoue2014} and a \cite{Kroupa2001} initial mass function. We perform SED modelling with an exponentially declining star-formation history, with a broad uniform prior in age from 0.1 to 9 Gyr and a uniform prior in ${\rm \tau}$, where ${\rm \tau \in (0.1, 9)\ Gyr}$. We set broad uniform priors on the stellar mass, where ${\rm \log M_\star / M_\odot \in (6, 13)}$, metallicity, where ${\rm Z/Z_\odot \in (0.003, 2)}$, ionisation parameter, where ${\rm \log U \in (-4, -1)}$ and we use a fixed velocity dispersion of ${\rm 100\ km/s}$. We use a \cite{Calzetti2000} dust law with a uniform prior in the range ${\rm A_V \in (0, 4)}$. The redshift is fixed at ${\rm z = 0.0195}$.

\begin{figure}
   \centering
   \includegraphics[width=0.45\textwidth]{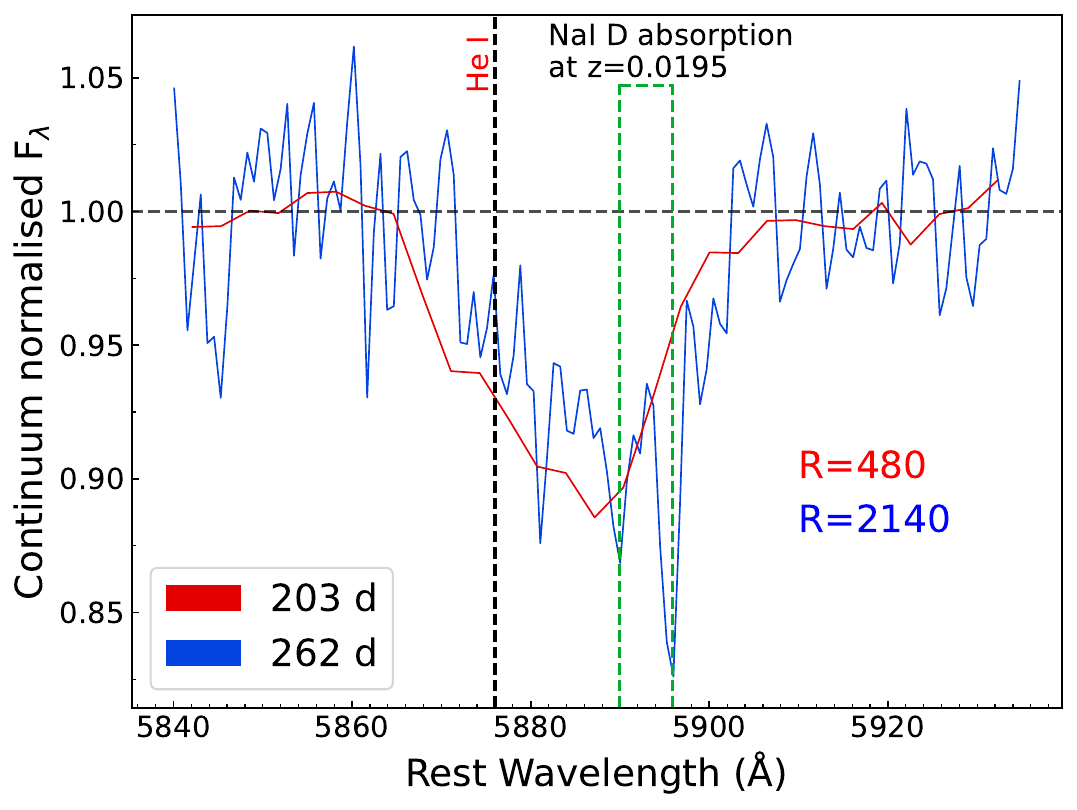} \newline
   \includegraphics[width=0.45\textwidth]{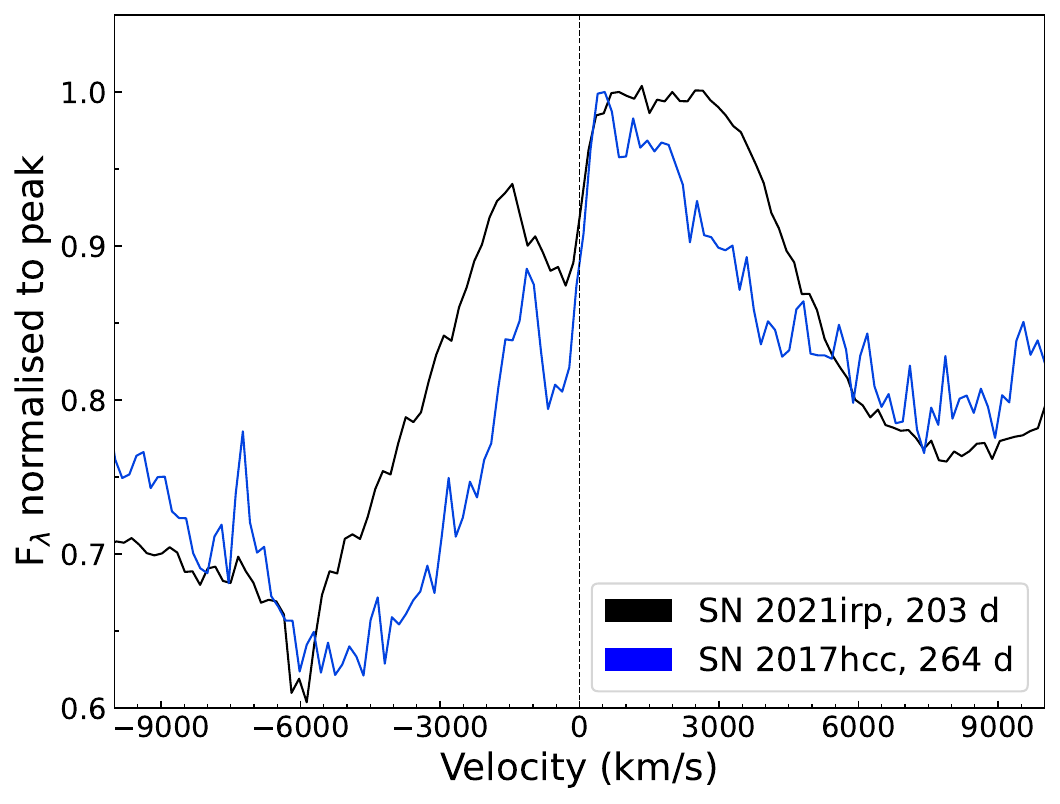} \newline
    \includegraphics[width=0.45\textwidth]{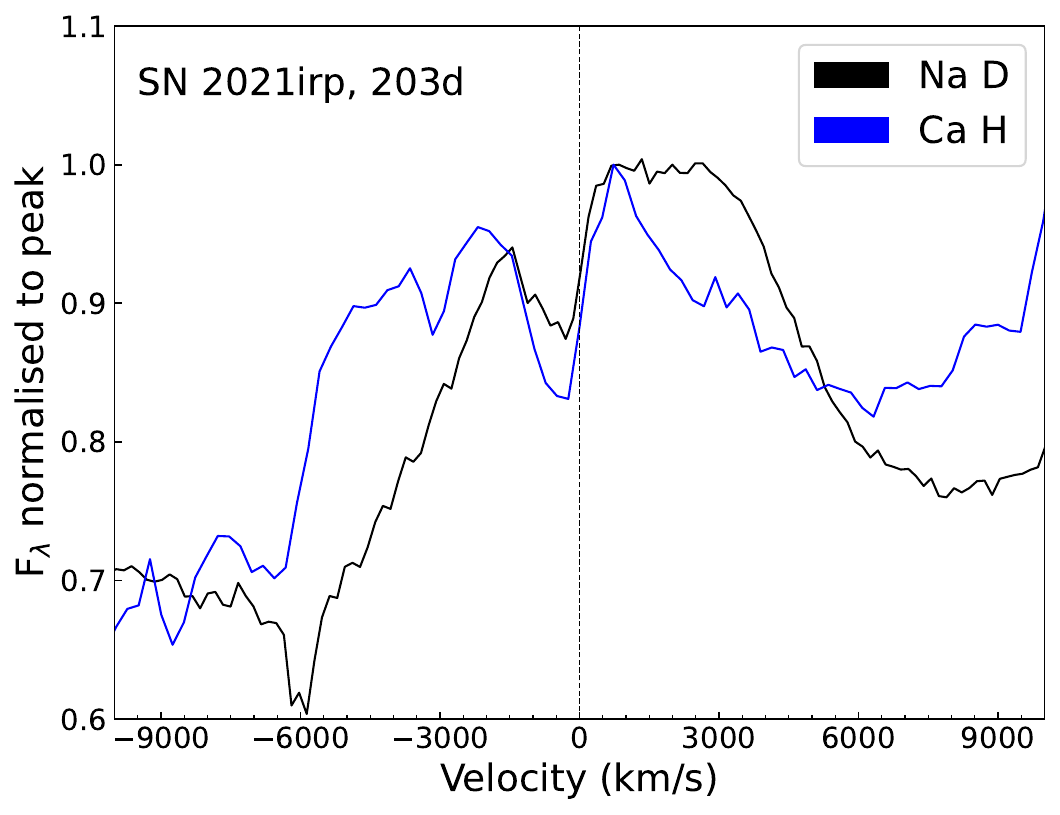} 

      \caption{\textbf{Top panel:} The \ion{Na}{I}~D absorption feature in our highest S/N spectra. In the +271~d spectrum, the individual narrow components of an absorption doublet at the host redshift can be seen. The resolving powers of the two spectra are listed. \textbf{Middle panel:} The \ion{He}{i}$\lambda$5876 / Na~{\sc i}~D emission feature for SN~2021irp and SN~2017hcc at a similar epoch, normalised at the emission line peak. Velocity is with respect to the midpoint of the Na doublet. Both SNe display an absorption feature superimposed in the emission profile. The narrower absorption at $\sim-6000$km~s$^{-1}$~is the MW Na absorption. \textbf{Bottom panel:} The absorption line in the spectrum of SN~2021irp associated with Ca H has a very similar width and blueshift to the Na D feature.}
\label{fig:Na_example}
\end{figure}

The host galaxy parameters inferred from our fitting are listed in Table \ref{tab:host_fitting_params}. The host of SN~2021irp is a typical star-forming galaxy, lying on the star-forming main sequence, which predicts a star formation rate (SFR) of $0.34\pm0.30~M_{\odot}$~yr$^{-1}$ \citep{Popesso2022}, consistent with our measured SFR of $0.51\pm0.03~M_{\odot}$~yr$^{-1}$. This is typical of host galaxies of Type IIn SNe or the general population of Type II SNe \citep{Schulze2021}. The metallicity of the host is one third Solar, between the metallicities of the LMC and SMC, which are half and a quarter Solar respectively \citep{Welty1999}. Compared to the LMC, which has a stellar mass of log($M_{*} / M_{\odot}$) = 9.4 \citep{vanderMarel2006} and a SFR of $\sim0.2~\text{M}_{\odot}$~yr$^{-1}$\citep{Harris2009}, the host of SN~2021irp has approximately half the stellar mass, and has $\sim$twice the SFR.

\subsection{Host extinction}
\label{sect:host_extinction}

An absorption feature at rest wavelength $\sim5885$~\AA~is visible in all the spectra of SN 2021irp with sufficient signal-to-noise. This lies close to the position of the \ion{Na}{I}~D ($\lambda\lambda$5889, 5895~\AA) doublet at the host galaxy redshift. In Fig. \ref{fig:Na_example}, we show a selection of the highest quality spectra where this is most clear. In particular, in our highest resolution spectrum, we can resolve the individual narrow components of the \ion{Na}{I}~D doublet at our adopted redshift of z = 0.0195, where these narrow features lie within a broader absorption feature. We assume that these resolved narrow features are the \ion{Na}{I}~D lines arising from obscuring material within the host galaxy in the line of sight.

On the other hand, the origin of the broad component is not straightforward. The broad absorption feature has a width of $\sim1000$~kms$^{-1}$, and the line centre is blueshifted by $\sim400$ kms$^{-1}$ compared to the centre of the \ion{Na}{I}~D doublet. If we identify this broad component with \ion{Na}{I}~D lines arising from obscuring material with different velocities \citep[see][for an example]{Kangas2016}, the interstellar gas should have velocities spreading from $\sim0$ to $\sim1400$ km/s. This is much larger than the typical velocity dispersion of gas clouds in a galaxy \citep{Green2010}, and larger than the expected rotational velocity of the galaxy \citep{Alves2000}, especially given that we observe the host of SN~2021irp from a face-on direction. In addition, the absorbing gas implied from this interpretation would consist of multiple clouds whose velocities are spreading continuously over $\sim0$ to $\sim1400$ km/s but not at negative velocities, which is unnatural. 

Crucially, SN~2017hcc, a Type IIn SN which is photometrically and spectroscopically similar SN to SN~2021irp (see Sects. \ref{subsec:photometric_comparison} and \ref{subsec:spec_comparison}), exhibits a similar broad \ion{Na}{I}~D absorption feature at similar epochs to SN~2021irp (see Fig. \ref{fig:Na_example}), but this feature is not present in spectra observed at early times \citep[see][]{Smith2020,Moran2023}. Therefore, we conclude the broad absorption feature visible in SN~2021irp to be a SN-related feature rather than an interstellar line, although we lack observations at early times to observe its appearance. 

In order to infer the host extinction, we measure the equivalent width (EW) of the two resolved narrow features of the \ion{Na}{I}~D doublet in our highest resolution spectrum. Fitting the continuum and 3 absorption lines (one broad, 2 narrow) yields a total EW only in the 2 narrow lines of $0.3\pm0.1$Å. Uncertainties are taken from the covariance matrix produced by the least square fitting procedure. Using the relationship of \citep{Poznanski2012}, this corresponds to $E(B-V)_{\text{host}} = 0.04\pm0.01$ or, adopting the Cardelli extinction law \citep{Cardelli1989} with $R_V = 3.1$, $A_{V,\text{host}} = 0.12\pm0.04$. This is close to the extinction of A$_{V} = 0.19^{+0.02}_{-0.01}$ we derived above from modelling the host galaxy SED, although this measurement is an average across the stellar populations of the host. We note that narrow absorption lines associated with the significant MW extinction are also detected and well resolved in the same spectrum, and measuring the equivalent width and converting to a E(B-V) with the \citet{Poznanski2012} relationship yields extinction values broadly consistent with those derived from the MW dust maps.

\section{\label{sec:PhotAnalysis}Photometric properties}

\begin{figure*}
   \centering
   \includegraphics[width=\textwidth]{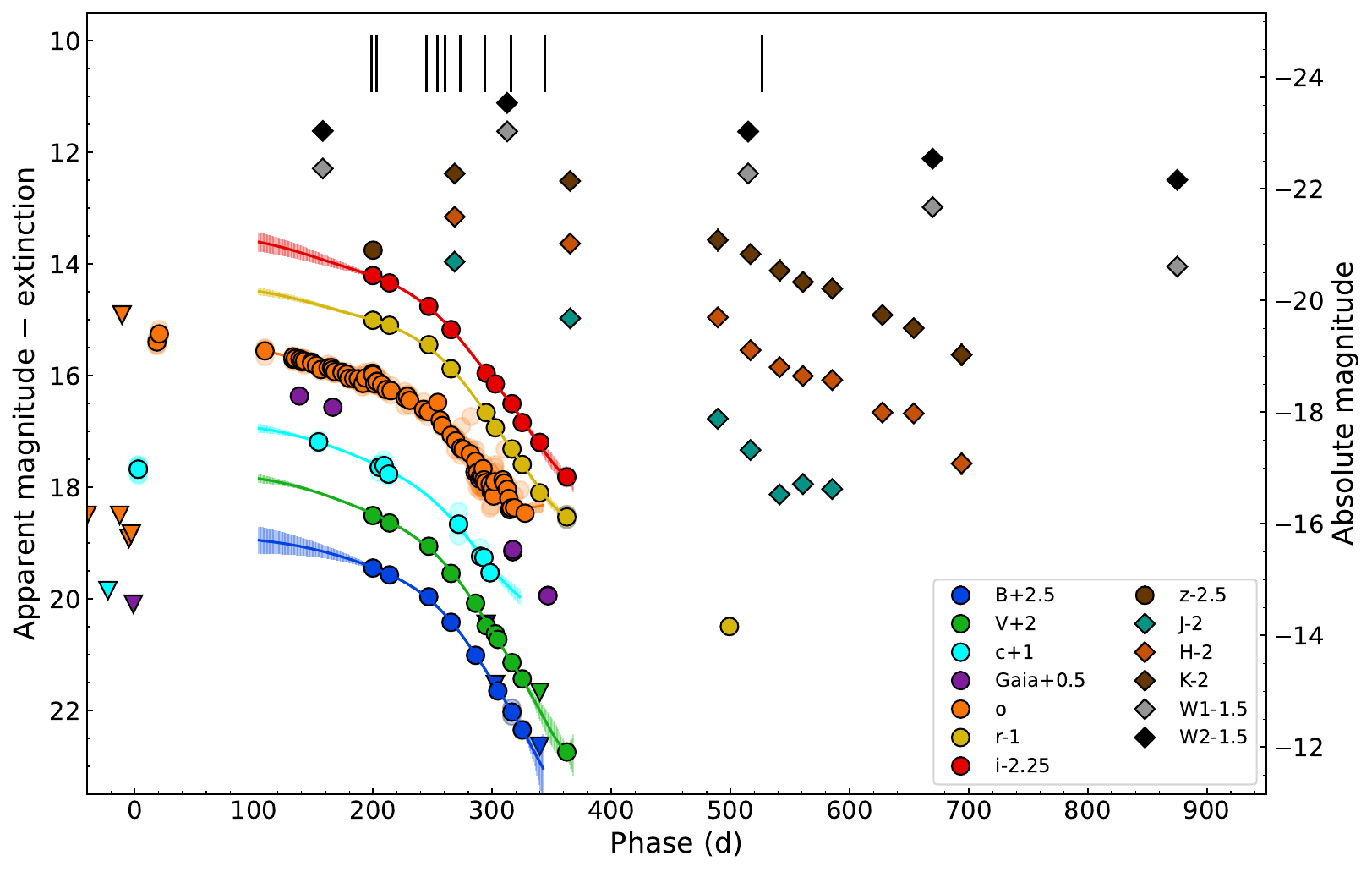}
      \caption{Optical and IR data of SN~2021irp. All data is corrected both for MW and host galaxy extinction. Vertical ticks represent the timing of spectral observations. Downward triangles are $3\sigma$ upper limits. Where there are multiple observations on the same night, the individual observations are shown with transparent symbols, and the binned measurement is shown with a solid symbol with a black outline. The solid lines show the results of Gaussian processing of the measured fluxes in both time and wavelength simultaneously, with the shaded region showing the associated uncertainties.}
\label{fig:LC}
\end{figure*}

\begin{figure}
   \centering
   \includegraphics[width=0.5\textwidth]{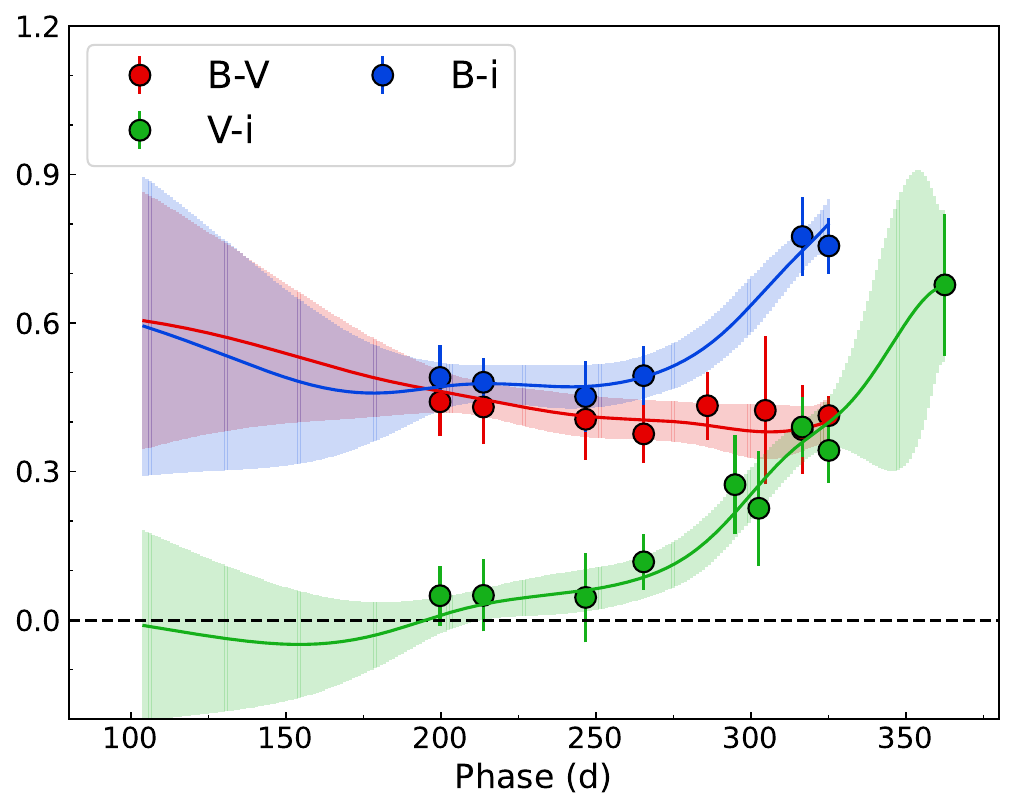} \newline
   \includegraphics[width=0.5\textwidth]{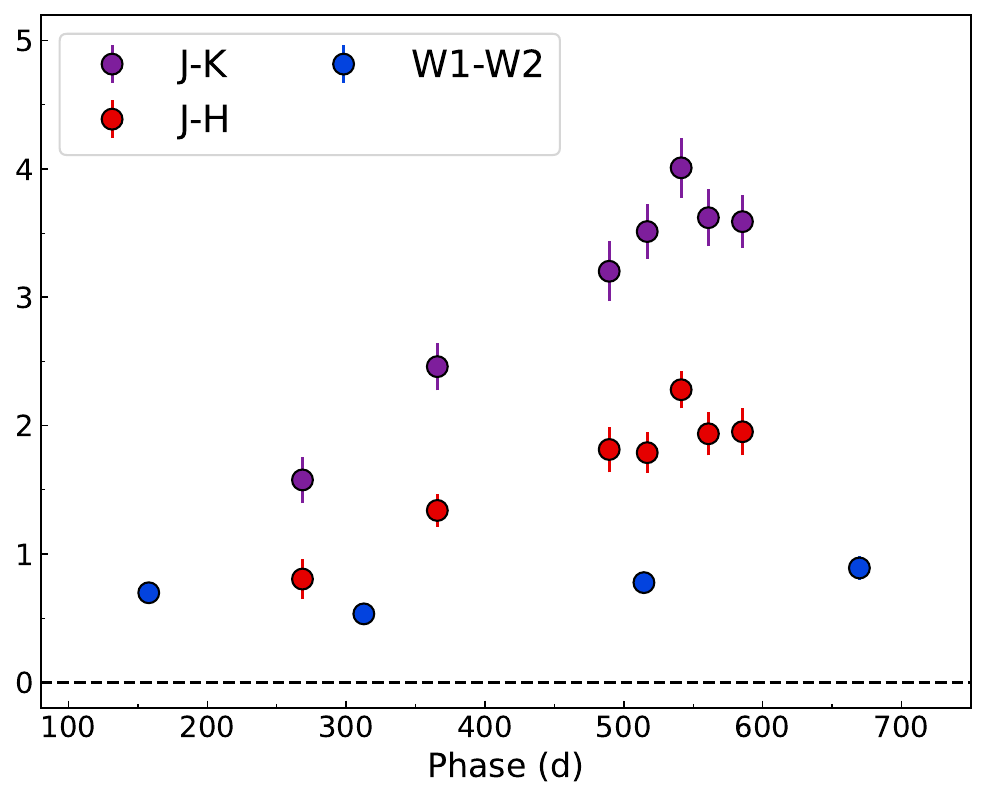}
      \caption{Colour evolution for SN~2021irp. \textbf{Upper panel:} Optical colour evolution. \textbf{Lower panel:} IR colour evolution.}
\label{fig:colors}
\end{figure}

We estimate the magnitude of the SN between observations with Gaussian processes. In particular, we used the functionality of the {\sc piscola}\footnote{\url{https://github.com/temuller/piscola}} package \citep{MullerBravo2022a} to perform a two-dimensional Gaussian process to interpolate the multi-colour observer frame light curves in time and wavelength simultaneously. Compared to fitting each band separately, this method produces very similar results for the regions where we densely sample the SED with photometry. However, this method also allows significant extrapolation of the $BVri$ light curves into the period between the first ATLAS $o$ observation after the SN returns from behind the sun at +111 d, until the beginning of our multi-band ALFOSC observations at +204 d, making use of the dense photometric time coverage in the ATLAS $o$ band. The single $c$ band epoch provides some constraint on the colour evolution during this period, and we assume that the Gaussian process well models the light curve and colour evolution. The photometry, along with the interpolated and extrapolated light curves, are presented in Fig. \ref{fig:LC}.

\subsection{\label{sec:LC+color}Light curve and colour evolution}

Due to the solar conjunction, we have relatively little information about the early phases of the evolution of SN~2021irp. The explosion epoch is well constrained within 3 days of our adopted value by the last ATLAS non-detection and the first ATLAS detection. The final observation before the conjunction occurred at 20.5~$\pm~3$~d when the SN had an absolute magnitude of $M_{o} = -19.4$~mag. There was an observation 2 days earlier where the SN was 0.15 mag fainter, indicating that it was likely still rising in our final observation. When the SN was next observed, in the $o$ band at 109.2~d, it had declined by 0.3 mag. Interacting SNe can exhibit long rise times, such as in the case of SN~2017hcc, which has a rise time of 57~d \citep{Moran2023}, or shorter rise times of $\sim20$~d \citep[see e.g.][]{Nyholm2020}. In the case of SN~2021irp, as it is likely that the peak magnitude is brighter than we observe, and the rise time is likely greater than the 20.5~$\pm~3$~d that we observed, but it is not possible to conclude more than that.

After returning from the solar conjunction, there is a linear decline in magnitude in the well-sampled $o$-band from 132~d until 230~d, with a slow decline rate of 0.0076 mag day$^{-1}$ derived from a linear fit to the data. The single $o$-band observation at 109~d lies below the linear fit, implying the decline rate was slower between 109~d and 132~d. After 230~d, the $o$-band decline rate increases, making a knee shape, before settling onto a much faster linear decline of 0.027 mag day$^{-1}$ between 250~d and 300~d. The expected decline rate from $^{56}$Co decay is 0.0098 mag day$^{-1}$ in the case of the full gamma-ray trapping \citep{Woosley1989,Miller2010}. SN~2021irp declines slower than this from 109~d until the break in the evolution at approximately 230~d, implying that $^{56}$Co decay is not the dominant power source. The slow decline followed by a relatively sharp increase in decline rate is reminiscent of the evolution of Type IIP SNe, which exhibit a plateau phase powered by the recombination of a massive H envelope, followed by a drop in brightness once the envelope is recombined \citep[e.g.][]{Arcavi2017}. However, the drop in the light curve of SN~2021irp occurs much later than that observed in typical Type IIP SNe, where the drop comes no later than $\sim135$~d \citep{Anderson2014,Takats2015}. Furthermore, we do not see the typical change of the spectral features from photospheric to nebular spectra that accompanies the drop in Type IIP SNe (see Sect. \ref{sec:spectra}).
This rapid luminosity drop is triggered by another mechanism than that in Type~IIP SNe. One possibility is that the decline rate of the energy input is intrinsically changed at $\sim230$ d. This would imply a sharp decrease in the energy input provided by a power source, most likely the CSM interaction. Alternatively, this might be explained by an apparent effect, where the optical light is being absorbed and re-radiated by freshly forming dust. We further discuss the origin of the rapid luminosity drop in Sect.~\ref{sec:discussion}.

In Fig. \ref{fig:colors}, we show the colour evolution of various filter pairs. The optical colour evolution is flat between 100~d and 250~d, 
although there are large uncertainties between 100~d and 200~d that reflect the very sparse SED coverage during this period. The colour evolution remains constant for the 3 epochs with good SED coverage between $\sim200$~d and $\sim250$~d. There is then a sharp turn to the red in both the $V-i$ and $B-i$ colours, beginning between the observations at 247~d and 265~d and continuing throughout the remaining observations. 

The MIR evolution of SN~2021irp shows a rise from the first detection by NEOWISE at 158~d to the peak detection at 313~d, followed by a decline until the final detection at 875~d. 
Our NIR coverage begins later than the first MIR detection, at 268~d, and we only observe the NIR light curve in decline. The decline rate between the first two observations in IR is significantly slower than that observed after the solar conjunction. The colour evolution in the IR is shown in Fig. \ref{fig:colors}. The colour consistently evolves to the red across all filters and all our time coverage, except for between the first two epochs of NEOWISE data, where the $W1-W2$ colour becomes slightly bluer. Both the sharp turn to the red in the optical colours and the IR colour evolution are likely due to dust formation. We discuss this possibility in Sect. \ref{sec:dust_formation}.

\subsection{\label{sec:BB_fits}Blackbody fitting and pseudo-bolometric luminosity}
 
\begin{figure}
   \centering
   \includegraphics[width=0.49\textwidth]{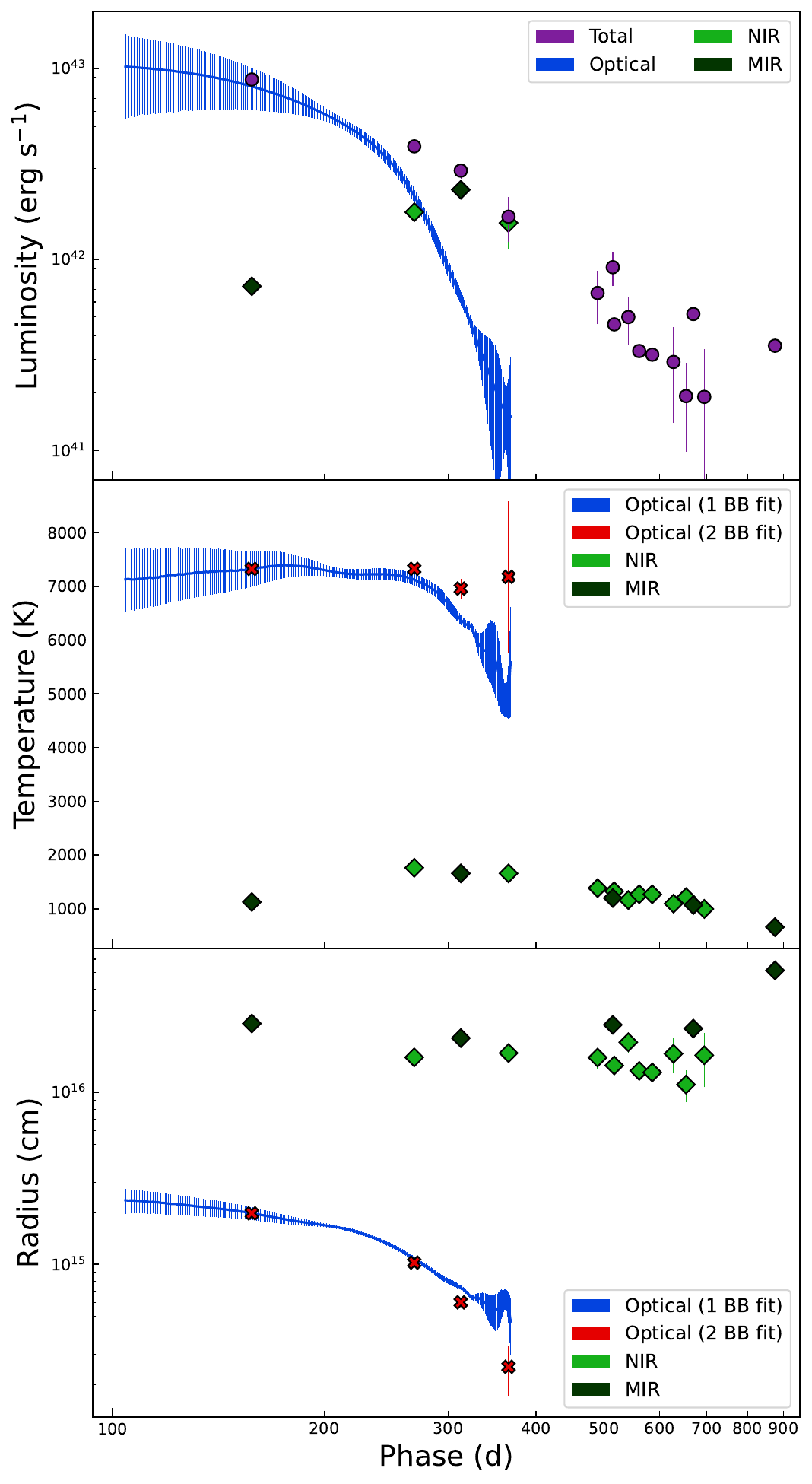}
      \caption{Results from blackbody fitting. The blue line indicates fitting to the interpolated optical light curves. The red crosses represent the fit parameters for the optical blackbody shown in Fig. \ref{fig:SED_fitting_IR}. The diamonds indicate parameters associated with the IR blackbody, with dark green indicating fitting to NEOWISE MIR photometry and lighter green fitting to NOTCam NIR photometry. \textbf{Top panel}: Pseudo-bolometric luminosities implied from the blackbody fitting. Luminosities were calculated from the Stefan-Boltzmann law. The total luminosity is the sum of the IR and optical luminosities. At later times where we have no optical data, the total luminosity is simply the IR luminosity. \textbf{Middle panel}: Evolution of the temperature for the optical and IR blackbodies. \textbf{Bottom panel}: Evolution of the radius for the optical and IR blackbodies.}
\label{fig:BB_bolo}
\end{figure}

\begin{figure}
   \centering
   \includegraphics[width=0.5\textwidth]{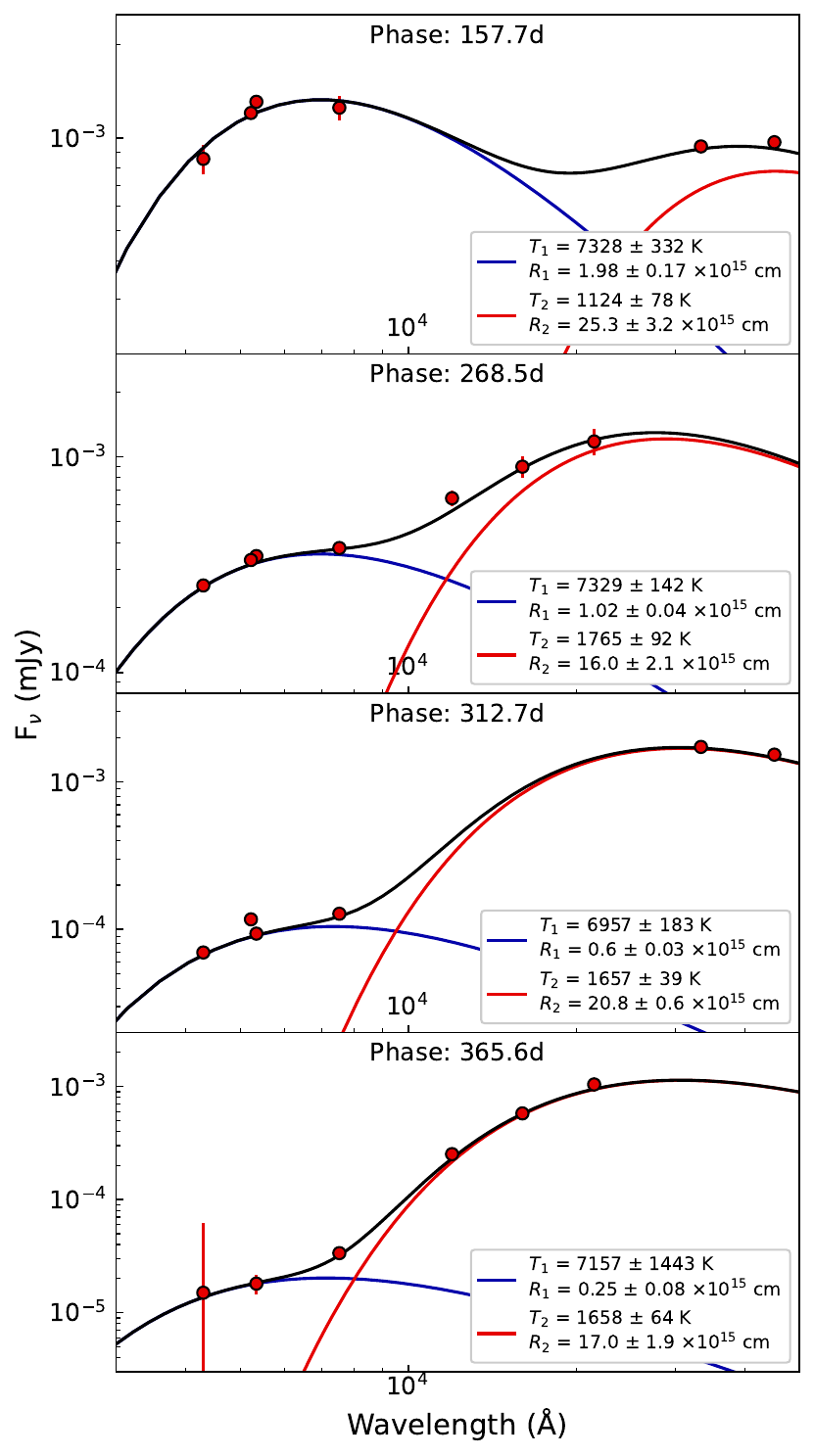}
      \caption{Two blackbody fitting for the epochs at which we can constrain both the optical and IR flux. Temperatures and radii for the SN and dust components are given in each panel.}
\label{fig:SED_fitting_IR}
\end{figure}

We construct a pseudo-bolometric light curve by fitting a blackbody to our photometry, using the {\sc emcee} python implementation of the Markov Chain Monte Carlo method \citep{ForemanMackey2013} to fit the points and derive uncertainties. Our photometric SED coverage in optical is dense from $B$ to $i$ band in the time period between 200~d and 350~d but is otherwise sparse. We make use of the results of the Gaussian processes driven extrapolation described above to obtain $BcVri$ light curves in the time period where we have primarily ATLAS $o$ data, as well as interpolate between our observations. We then construct the pseudo-bolometric light curve using the $BcVi$ bands, as the strong H$\alpha$ emission line falls in the $r$ and $o$ bands. After obtaining an estimate of the temperature and radius, we calculate the associated luminosity with the Stefan-Boltzmann law. The parameters derived from this fitting are shown in Fig. \ref{fig:BB_bolo}. At the epochs where we have IR data, we construct the total SED from the interpolated optical light curves and the IR observations and fit simultaneously with two blackbodies to derive temperatures, radii and total luminosities. These SED fits, shown in Fig. \ref{fig:SED_fitting_IR}, yield good fits to a blackbody, except for an excess in $J$ band at 269~d, which we attribute to the likely presence of line emission from Pa$\gamma$, Pa$\beta$ and \ion{He}{I} $\lambda$10830, which are all observed to be strong in similar SNe such as SN~2017hcc and SN~2010jl at similar epochs. \citep{Moran2023}.

Similarly to the optical photometry, the optical blackbody luminosity slowly declines with a relatively constant rate between 100~d and 200~d before the decline rate increases at $\sim230$~d, forming a knee shape. The decline rate then becomes constant again, at a faster rate, just before 300d and continues on this decline until our final observations. The temperature is quite constant throughout the 100~d - 250~d period at $\sim6000$ - $7000$ K (although there is large uncertainty on the temperature before 200d) before sharply declining between the observations at 265~d and 285~d. This temperature is consistent with those of Type II SNe during the photospheric phase, matching with the hydrogen recombination temperature in their ejecta \citep[e.g.][]{Faran2018,Martinez2022}. The photospheric radius is initially a few $\times 10^{15}$ cm, and declines by an order of magnitude by our final observation, with the shape of the evolution following the luminosity decline relatively closely. As the optical blackbody luminosity declines, the IR luminosity increases to a measured peak at 313~d.

If the IR emission arises from optical emission reprocessed by newly formed dust (see Sect.~\ref{sec:dust_formation} for the discussion on this possibility), then the measured temperatures and radii for the SN will be affected. The emitted flux in the optical will be larger than the observed flux, leading to an underestimate of the photospheric radii and luminosity. This is consistent with the change in decline rate that we observe during the period from $\sim200 - 300$~d, during which the ratio of luminosities $L_{\text{dust}}/L_{\text{SN}}$ increases from $\sim 0.1$ to $\sim 3$. Therefore, the photospheric radii during this period and afterwards are likely underestimated. This effect also leads to underestimating the temperature derived from single blackbody fits due to preferential extinction in the bluer bands and emission red optical bands, both effects due to dust. As shown in Fig. \ref{fig:BB_bolo}, the optical blackbody temperature derived from 2 blackbody fitting (Fig. \ref{fig:SED_fitting_IR}) does not show any decline in the 268~d - 365~d period as the contribution in the $i$ band from the second, cool blackbody is included in the fit.

Our first IR observation at 158~d yields a much lower luminosity and temperature and a larger radius than the second at 269~d. This first measurement arises from only 2 MIR points, which are not optimal for observing hot dust, as the blackbody peaks at shorter wavelengths at these temperatures, e.g., a 2000~K blackbody peaks at 1.45~$\mu$m, in $H$ band. However, the intrinsic MIR colour is red, with the 3.4~$\mu$m flux density being less than that at 4.6~$\mu$m. Furthermore, there is a significant contribution from the SN at 3.4~$\mu$m, implying that the IR colour due only to dust emission is even redder. Therefore, we consider the cooler temperature to be accurate. This temperature and radius evolution for the dust can be explained if the IR emission in this first measurement arises due to an IR echo from distant pre-existing dust, and we discuss this possibility in Sect. \ref{sec:dust_formation}.

The IR luminosity declines relatively smoothly from the peak until the final observation at 874~d. The temperature evolution is consistent with radiation from dust, with the maximum measured temperature of $1680 \pm 60$ K being less than the sublimation temperature of graphite grains of $\sim2000$ K, although this is likely too hot for the survival of silicates which are destroyed at these temperatures \citep[see e.g.][]{Guhathakurta1989}. Excluding the first observation, the temperature declines smoothly from its peak while the radius stays quite constant at $1.5 - 2\times~10^{16}$~cm.

By integrating the luminosity measurements, we can estimate the total radiated energy during our observations to be $9.9\times10^{49}$ erg in the optical, and $7.2\times10^{49}$ erg in the IR, or $1.7\times10^{50}$ erg in total. Notably, this estimate excludes the first 109~d of the SN when the SN was behind the sun, during which the optical luminosity is expected to almost always be larger than the largest values we observe. Conservatively, if we estimate the unobserved luminosity as simply the largest estimated optical pseudo-bolometric luminosity we observe multiplied by the time the SN went unobserved, we expect to have missed another $9.1\times10^{49}$ erg, yielding an estimate for the total radiated energy of $2.6\times10^{50}$ erg. This is $\sim 10\%$ of the maximum available explosion energy available for a neutrino-driven core-collapse SN \citep{Janka2016}.

\subsection{\label{subsec:photometric_comparison}Photometric comparison}

\begin{figure}
   \centering
   \includegraphics[width=0.5\textwidth]{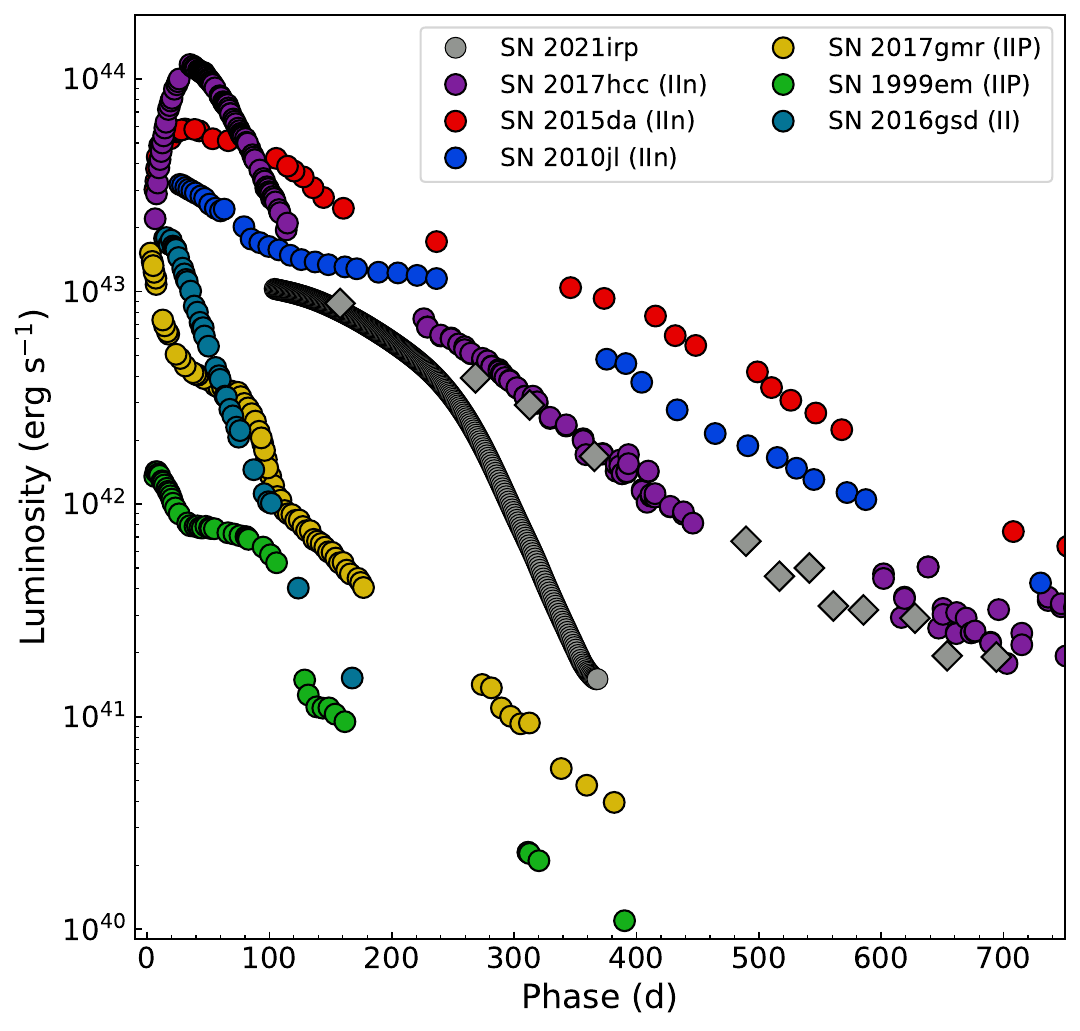}
      \caption{Pseudo-bolometric light curve for SN~2021irp compared with those of other SNe. The grey diamonds show the total pseudo-bolometric luminosity at epochs where we have IR observations. Data references are given in Table \ref{tab:comparison_objects}}
\label{fig:bolo_comp}
\end{figure}

In Fig. \ref{fig:bolo_comp}, we compare the pseudo-bolometric light curve for SN~2021irp with other selected SNe from the literature. We compare with a selected sample of well-observed luminous and long-lasting Type IIn SNe, namely SN~2010jl, SN~2017hcc, and SN~2015da, which we selected based on spectral similarity to SN~2021irp (see \ref{subsec:spec_comparison}). We additionally compare with the luminous Type IIL SN~2016gsd
; and some Type IIP SNe: SN~1999em and SN~2017gmr. The references for the data and adopted parameters for these SNe are listed in Table \ref{tab:comparison_objects}. All the pseudo-bolometric light curves were constructed in the UV+optical+IR wavelength range, either directly from observations or estimating the contribution in the UV and IR from the optical blackbody. The light curves for SN~2010jl and SN 2015da were constructed through direct integration including the NIR region, and include some contribution of the IR excess arising from dust emission (see \citet{Fransson2014} and \citet{Tartaglia2020} for details)

When it is first observed after the solar conjunction at 109~d, SN~2021irp is much more luminous than typical Type II SNe, either Type IIP or luminous Type IIL SNe. Furthermore, it remains at such a high luminosity for an additional 100 days or more, and even after a steep decline in luminosity, it does not reach the low luminosity levels of the Type II SNe in their radioactive powered tail phase. This demonstrates that an additional energy source is required to power the luminosity of SN~2021irp compared to the typical power sources in Type II SNe, namely radioactive power and the release of thermal power from the SN envelope.

\begin{figure*}
   \centering
   \includegraphics[width=\textwidth]{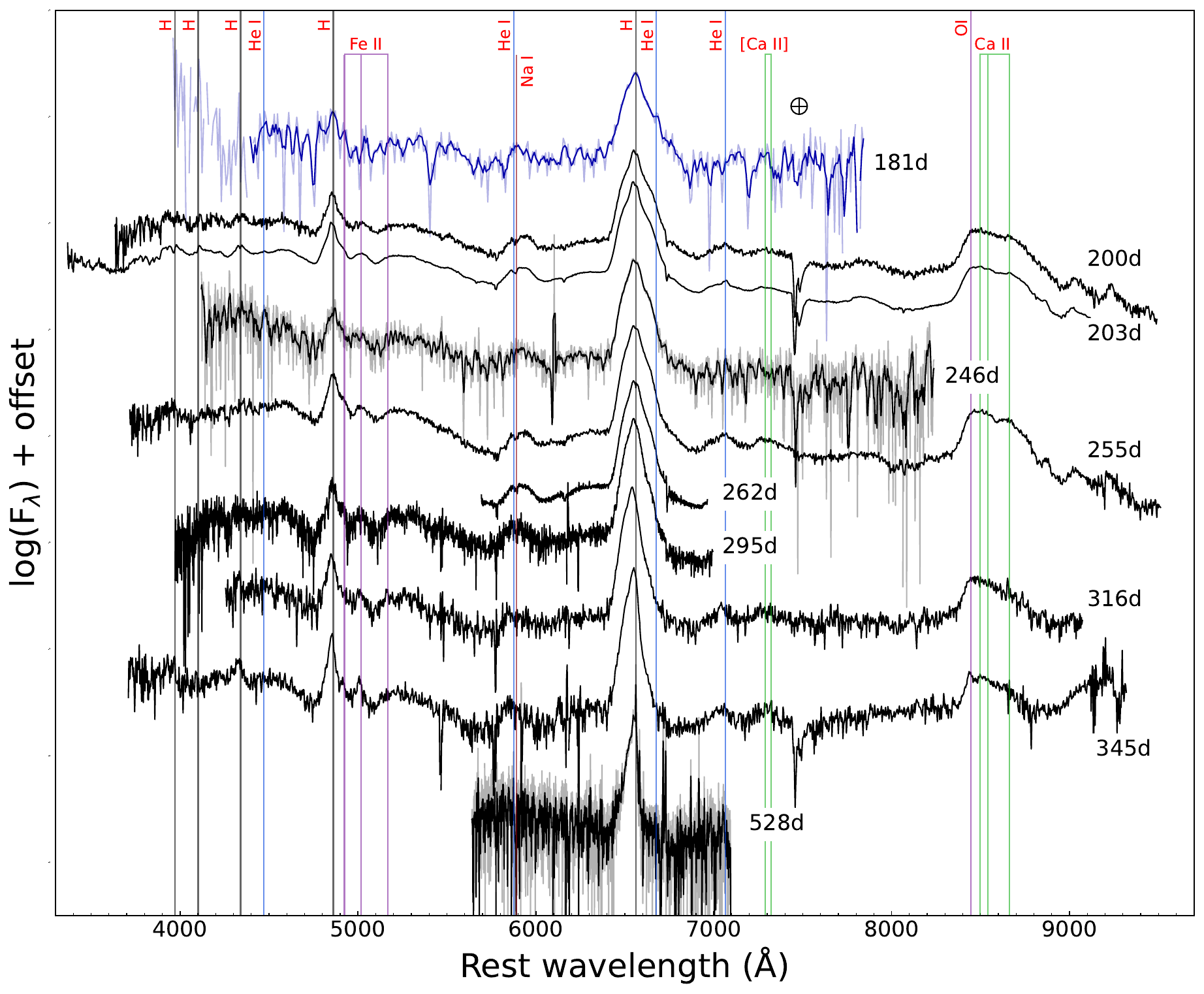}
      \caption{Spectra of SN~2021irp. All spectra are corrected for host and MW extinction. The cross-hair symbol denotes the region strongly affected by telluric absorption, which is corrected in some spectra.}
\label{fig:spectra}
\end{figure*}

Our comparison luminous and long-lived Type IIn SNe all show extensive interaction, with a few - 10 M$_{\odot}$ of CSM being inferred from observations and modelling \citep[see e.g.][]{Fransson2014,Tartaglia2020,Smith2020}. SN~2021irp is relatively similar in luminosity to SN~2017hcc and SN~2010jl in the 100d - 250d period, being a factor of a few less luminous, suggesting that similarly large masses of CSM could be required to power the SN during this period. There appears to be an increase in the decline rate for both SN~2017hcc and SN~2010jl at a similar epoch as SN~2021irp, although it is less pronounced. Both these SNe presented evidence of dust formation \citep{Smith2020,Sarangi2018,Gall2014}, which could point to this decline being related to dust formation in all three SNe, with different extents and timings. However, interacting SNe light curves exhibit great diversity due to the differing CSM properties, which can also explain these changes in decline rate. Due to the steep decline, by the final observations at $\sim400$~d SN~2021irp is more than an order of magnitude less luminous than SN~2017hcc and SN~2010jl. However, the total luminosity, including the IR component, is very similar in luminosity to SN~2017hcc for the entire period from 300d until 700d. We note that the significant luminosity arising from dust is not included in the pseudo bolometric light curve of SN~2017hcc, although it is 50\% of the total at +399d.

This comparison sample includes only luminous and long-lived interacting SNe, whereas the majority of Type IIn SNe are much less extreme. The average peak $r$-band magnitude for the Type IIn in the sample of \citet{Nyholm2020} (corrected for Malmqvist bias) was $\sim-18.6$, significantly less luminous than SN~2021irp. The systematic study of \citet{Hiramatsu2024} found the population of Type IIn SNe to be bimodal, divided into the  faint-fast and luminous-slow groups. SN~2021irp falls distinctly into the latter group.

As we almost entirely lack observations of SN~2021irp from 0 to 132~d, we can only compare to similar objects in order to speculate about the behaviour at peak. SN~2017hcc is almost an order of magnitude more luminous at peak compared to 100~d, while SN~2010jl is only about twice as luminous. This large variation prevents much meaningful speculation about the behaviour of SN~2021irp during the period without observations.

\section{\label{sec:spectra}Spectroscopic properties}

\subsection{\label{subsec:spectra_evolution}Spectral evolution}

We show the spectral evolution of SN~2021irp in Fig. \ref{fig:spectra}. The first spectrum was not observed until 2021-10-08, 182~d after the explosion, and shows only broad H$\alpha$ and H$\beta$ lines due to poor data quality. Our first follow-up spectra at 200~d and 203~d show broad emission lines of H$\alpha$, H$\beta$, H$\gamma$ and H$\delta$; \ion{He}{I} $\lambda$5876~\AA~and \ion{He}{I} $\lambda$7065~\AA; the Ca NIR triplet \ion{Ca}{II} $\lambda\lambda\lambda$8498, 8542, 8662~\AA; and forbidden [\ion{Ca}{II}] $\lambda$7291, 7323~\AA. The feature at $\sim5876$~\AA~is likely to have a significant contribution from both \ion{He}{I} $\lambda$5876~\AA~, as we also observe the \ion{He}{I} $\lambda$7065~\AA~line, and \ion{Na}{I}~D, as the feature is slightly redshifted and \ion{Na}{I}~D is more typical to observe at this epoch in Type II supernovae. There are no clear, strong, broad absorption lines or P-Cygni profiles. There appears to be a P-Cygni profile associated with the \ion{He}{I} $\lambda$5876 / \ion{Na}{I}~D feature, but this is more likely due to an \ion{Fe}{II} pseudo-continuum that rises at approximately 5600~\AA, which is seen in a number of interacting SNe \citep[e.g. 2005ip, see][]{Stritzinger2012}.

\begin{figure}
   \centering
   \includegraphics[width=0.5\textwidth]{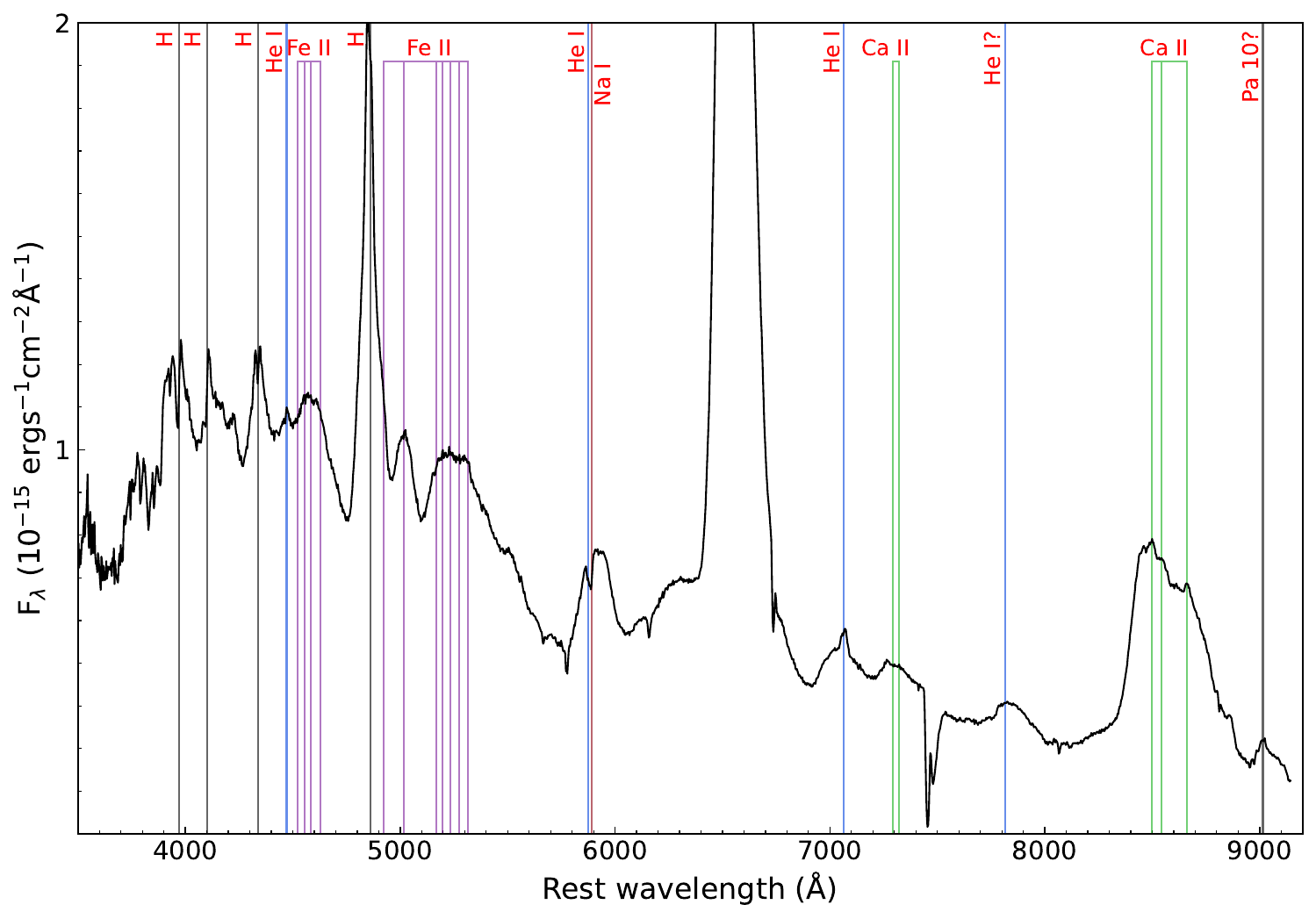}
      \caption{Spectrum taken at 203 d, displayed to show the continuum and less strong emission lines.}
\label{fig:spec_zoom}
\end{figure}

In Fig. \ref{fig:spec_zoom}, we show the spectrum at 203~d to highlight the continuum shape and the weaker features. There are broad bumps in the spectra corresponding to blended \ion{Fe}{II} emission lines at $\sim4600$~\AA~and 5200~\AA. There is a weak, narrow feature at 4471~\AA~that we associate with \ion{He}{I} $\lambda$4471.5~\AA. There is a broad feature at $\sim7850$~\AA~ that we can not definitively identify. \ion{O}{I} $\lambda$7772~\AA~would be very redshifted, and \ion{He}{I} $\lambda$7816~\AA~would be surprising given the relative strength of this line, as well as being offset from the line centre. There is a feature at $\sim9018$~\AA~that we tentatively identify with Paschen 10-3, as observed in e.g. SN~2010jl at late times \citep{Fransson2014}. Narrow absorption lines corresponding to \ion{Na}{I}~D absorption and diffuse interstellar bands \citep[DIBs; see e.g.][]{Herbig1995,Geballe2016} are visible at zero redshift, arising due to the significant extinction from the MW.

As discussed in Sect. \ref{sect:host_extinction}, there is an absorption feature with width $\sim1000$~km~s$^{-1}$ within the broad emission feature associated with \ion{He}{I} $\lambda$5876 / \ion{Na}{I}~D that is visible in all the spectra of SN~2021irp with sufficient signal to noise. We demonstrate that this feature is associated with the transient and not the host galaxy in Sect. \ref{sec:host_galaxy}. The intermediate width feature could be associated with either \ion{He}{I} $\lambda$5876~\AA~or \ion{Na}{I}~D. We see an intermediate width \ion{He}{I} $\lambda$7065~\AA~emission feature at the same epoch, with a larger width of $\sim2000$~km~s$^{-1}$. It is unlikely that we would see such a combination of emission and absorption, and furthermore, the absorption feature would be redshifted by $\sim500$~km~s$^{-1}$ if associated with \ion{He}{I}. These properties cannot be explained by normal SN-ejecta-originated lines, i.e., absorption lines created in homologously expanding optically thick gas. We therefore suggest that the feature is more likely arising from \ion{Na}{I}~D. In this case, the line would be blueshifted by $\sim400$~km~s$^{-1}$.  Interestingly, there are also absorption lines associated with \ion{Ca}{II} H and K (3968~\AA~and 3934~\AA) with very similar velocities observable in the same spectra, shown in Fig. \ref{fig:Na_example}, which may support this interpretation. This velocity is too low for the line to plausibly arise in the SN ejecta, as such velocities correspond to locations in the very central regions of the ejecta. The line should therefore arise in the CSM. One possibility is that the line is associated with CSM that was explosively ejected from the progenitor at this relatively high velocity. However, as described below, we constrain the velocity of the unshocked CSM for SN~2021irp to < 85 km~s$^{-1}$, so we would require a CSM with multiple components moving at different velocities simultaneously. 
One possibility 
is that initially slow moving nearby CSM was ionised and accelerated by the strong SN radiation, as an extreme case of what is observed in some Type Ia-CSM SNe \citep[e.g.,][]{Patat2007}, and is consistent with some theoretical work \citep{Tsuna2023}.

\begin{figure}
   \centering
   \includegraphics[width=0.5\textwidth]{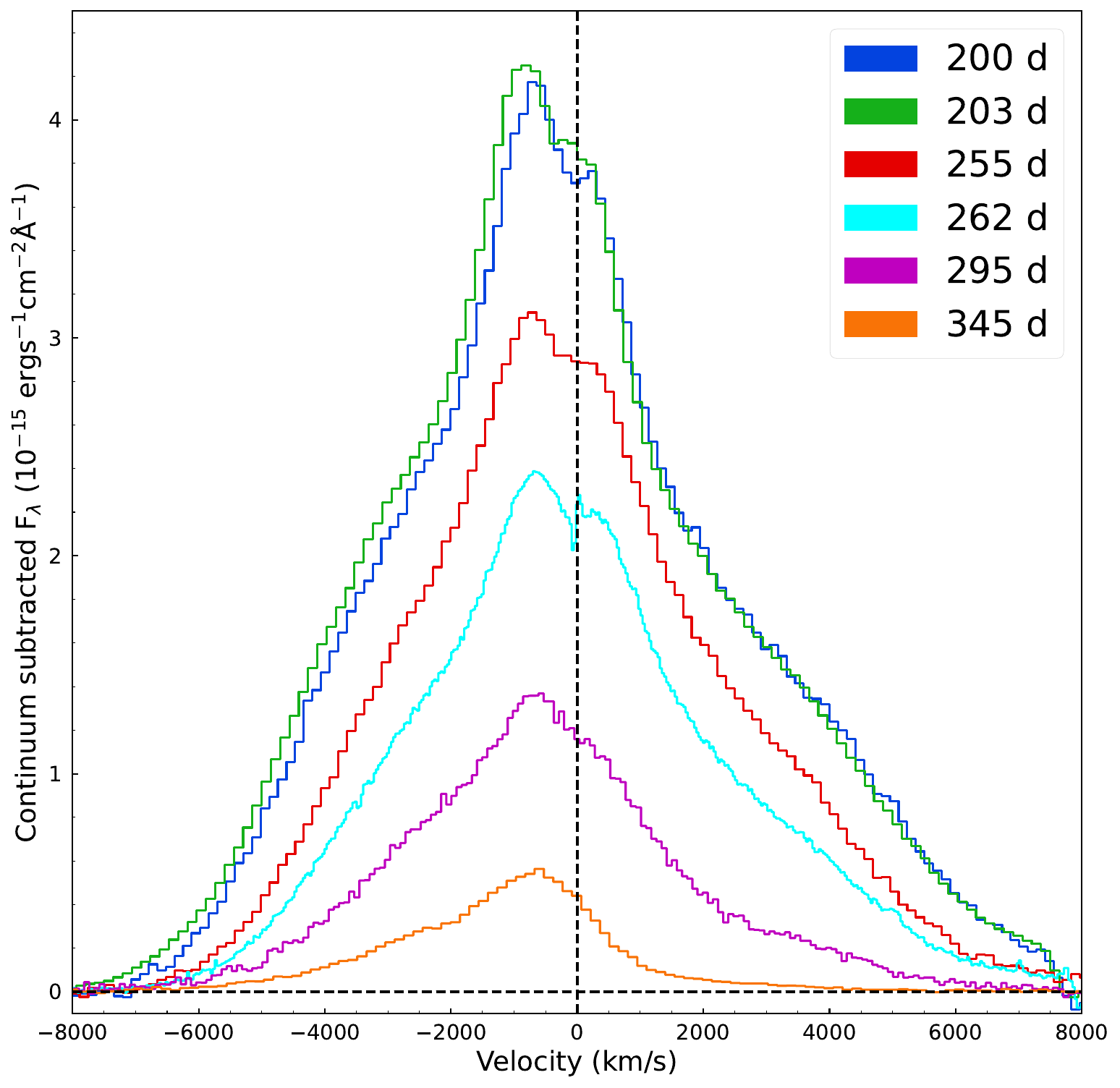}
   \includegraphics[width=0.5\textwidth]{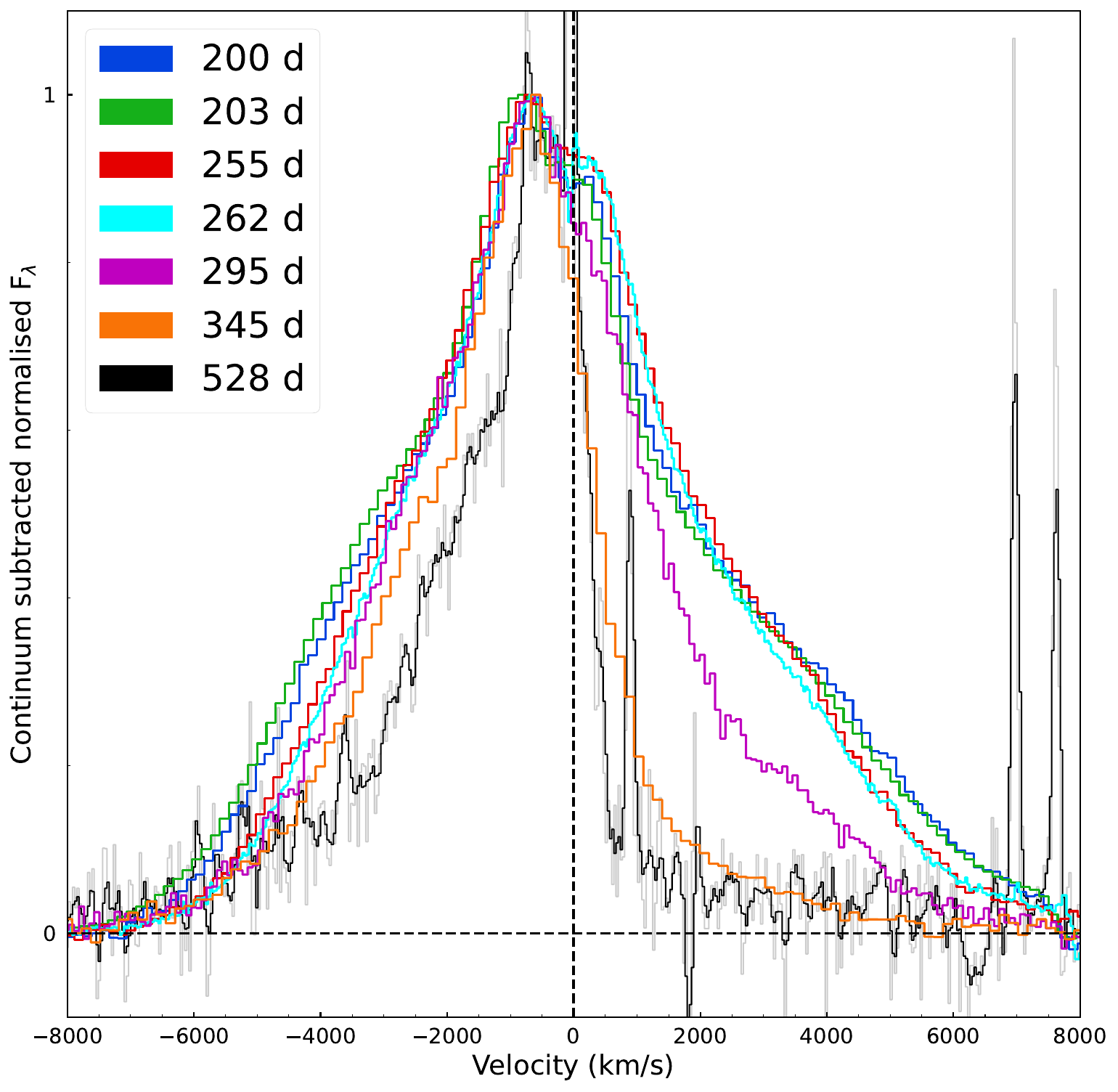} \newline
      \caption{The evolution of the H$\alpha$ line. \textbf{Upper panel:} Spectra are continuum subtracted. \textbf{Lower panel:} Spectra are continuum subtracted and normalised to peak at 1. The spectrum at 528~d has been smoothed for visual clarity, the un-smoothed spectrum is shown in the light grey. This spectrum was normalised ar the peak of the broad feature, between the narrow H$\alpha$ and [\ion{N}{II}] lines.}
\label{fig:Halpha comp}
\end{figure}

\begin{figure}
   \centering
   \includegraphics[width=0.5\textwidth]{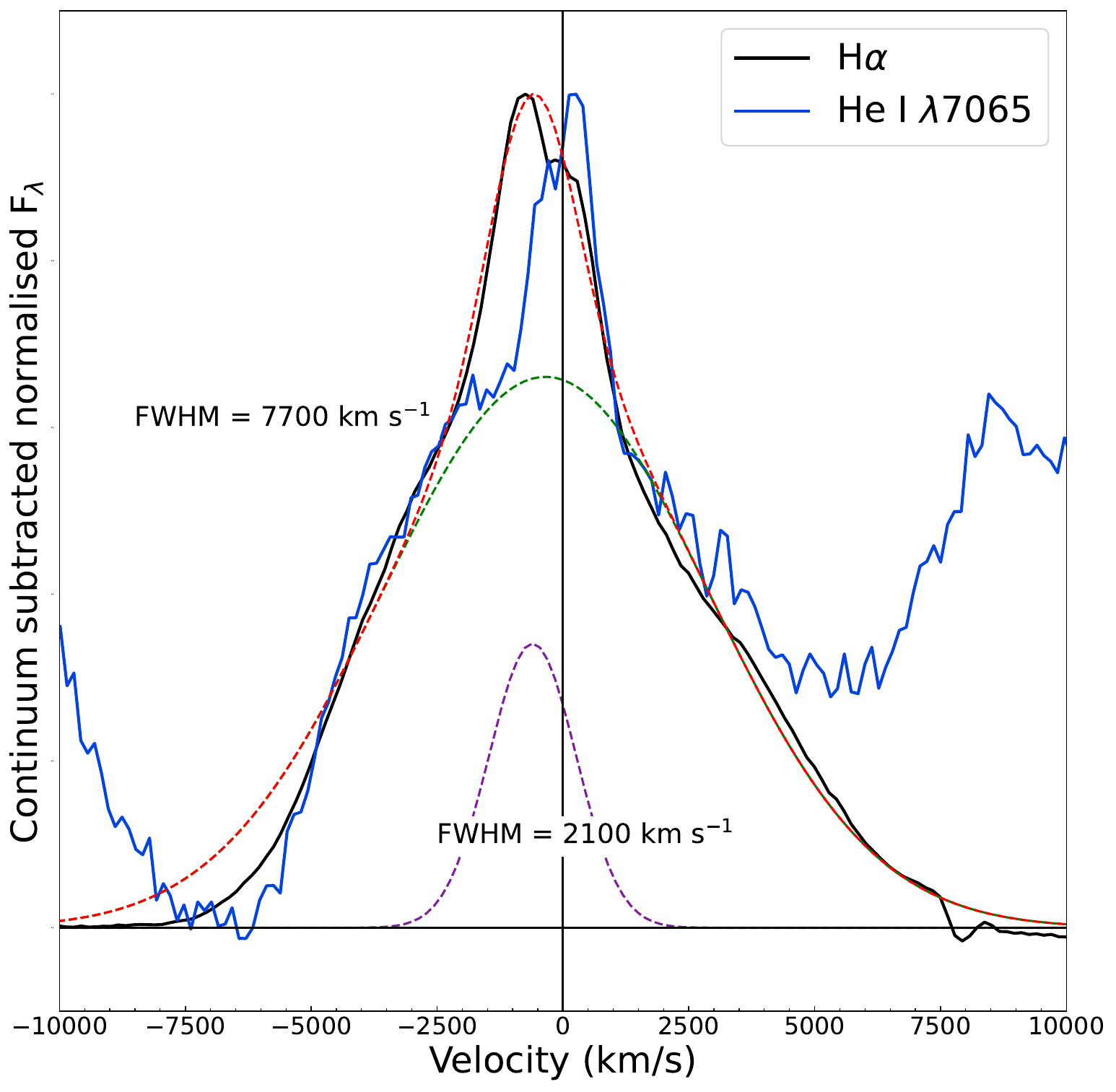}
      \caption{H$\alpha$ line profile compared to that of \ion{He}{I} $\lambda$7065 at 203 d. The continuum has been subtracted and the peak of the emission feature normalised to unity. Additionally shown are Gaussian profiles fit to the H$\alpha$ profile, with their FWHM listed. }
\label{fig:Ha_HeI}
\end{figure}

The emission line profiles are complex, and in particular, the Balmer lines exhibit a multi-component profile, shown in Fig. \ref{fig:Halpha comp}. Fitting two Gaussians to the line profile of H$\alpha$ at 203~d yields a broad and an intermediate component with FWHM of $\sim7700$ km~s$^{-1}$ and $\sim2100$ km~s$^{-1}$ respectively, which are shown in Fig. \ref{fig:Ha_HeI}, but these fits do not capture the complexity of the line profile and we only list these velocities to give an approximate indication of the width of the lines. The intermediate component is asymmetric, with the peak of the line profile being offset by $\sim-600$~km~s$^{-1}$ from the rest wavelength and exhibiting an uneven profile, with a deficit of emission on the red side. H$\beta$ also exhibits a multi-component profile, as does the \ion{He}{I} $\lambda$7065 line at 203~d, which is shown in Fig. \ref{fig:Ha_HeI}. The line profile of \ion{He}{I} $\lambda$7065 is similar to that of H$\alpha$ at the same epoch, except the intermediate width component is a reflected counterpart to that of H$\alpha$, with a redshifted peak for the emission feature rather than blueshifted, and is narrower.

\begin{figure}
   \centering
   \includegraphics[width=0.5\textwidth]{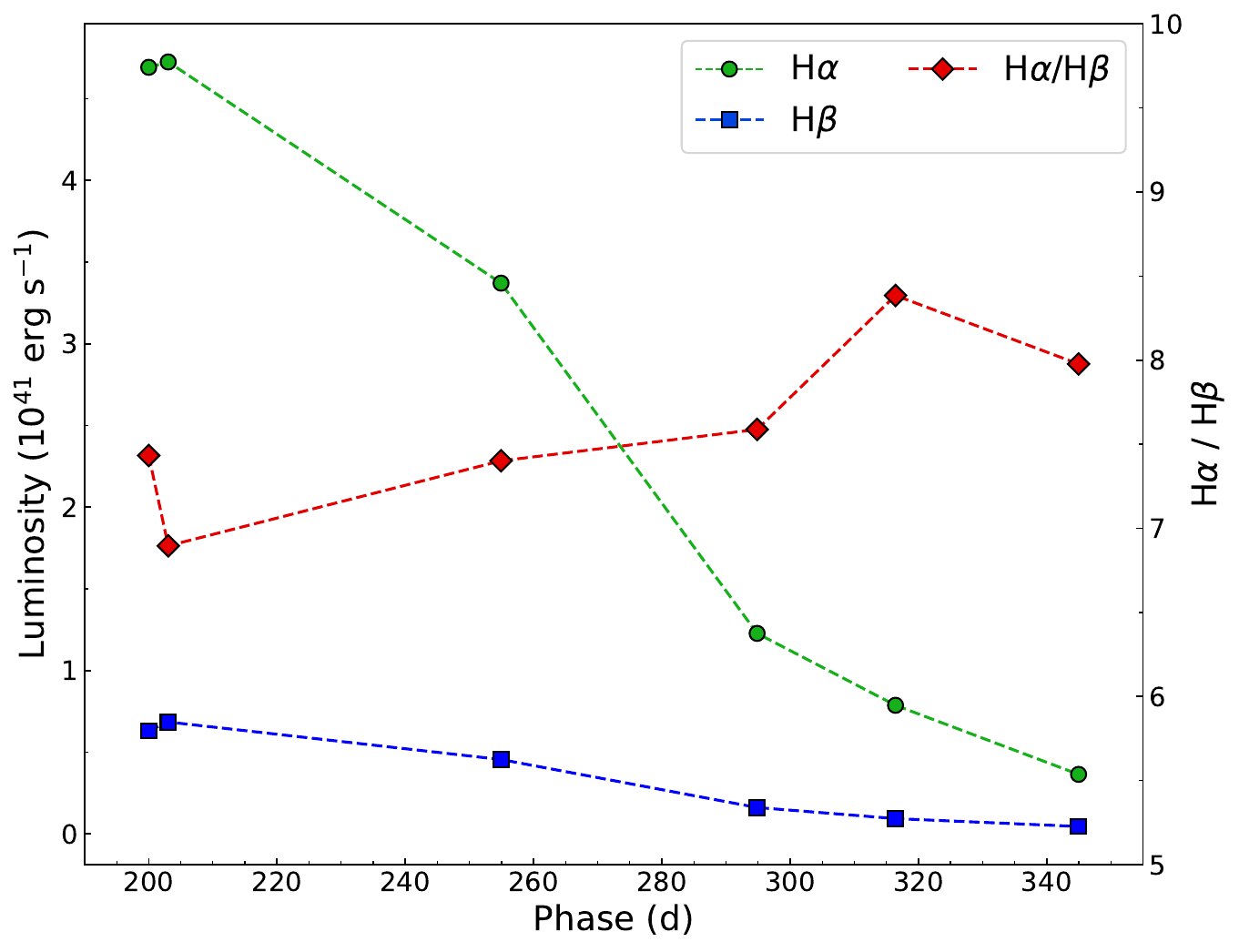}
      \caption{Line luminosities for H$\alpha$ and H$\beta$ as well as their ratio. Measurement uncertainties are smaller than the symbols.}
\label{fig:Balmer_measurements}
\end{figure}

To measure the flux for the H$\alpha$ and H$\beta$ emission lines we first estimate and subtract the continuum by performing a linear fit to selected regions on both the blue and red side of the emission line. For H$\alpha$, the region around the line is smooth and we use the $10000$ - $8500$ km~s$^{-1}$ region on both the blue and red side. For H$\beta$, the continuum is less clear, due to the presence of a number of Fe emission lines. On the blue side, we use the $-9000$ - $-5500$ km~s$^{-1}$ region, corresponding to the local minimum of the spectrum between the H$\beta$ emission line and the \ion{Fe}{II} blended emission bump. On the red side, we use the $+5500$ - $+8500$ km~s$^{-1}$ region, corresponding to the minimum between the H$\beta$ emission line and the \ion{Fe}{II}~$\lambda$5018. We note that there are significant systematic uncertainties in this continuum measurement, and that the broad H$\alpha$ and H$\beta$ lines could be blended with \ion{He}{I} $\lambda$6678 and \ion{Fe}{II} $\lambda$4924 respectively. We measure the line flux in the continuum subtracted spectrum directly, making use of the tools available in the {\sc specutils}\footnote{\url{https://github.com/astropy/specutils}} software package. The results are shown in Fig. \ref{fig:Balmer_measurements}. The Balmer decrement is $\sim7$ in our first spectral observations, and grows slightly over the course of our observations. This is much larger than the Case B value of $\sim3$ \citep{Osterbrock2006}, which could be explained by high densities \citep{Drake1980}, and is an indirect sign of strong CSM interaction. Similarly high values have been observed in other Type IIn SNe such as SN~2010jl \citep{Fransson2014}. Alternatively, absorption by dust can lead to a high Balmer decrement, and our IR observations indicate that dust is present from at least 158~d, before our earliest spectrum. 

\begin{figure}
   \centering
   \includegraphics[width=0.5\textwidth]{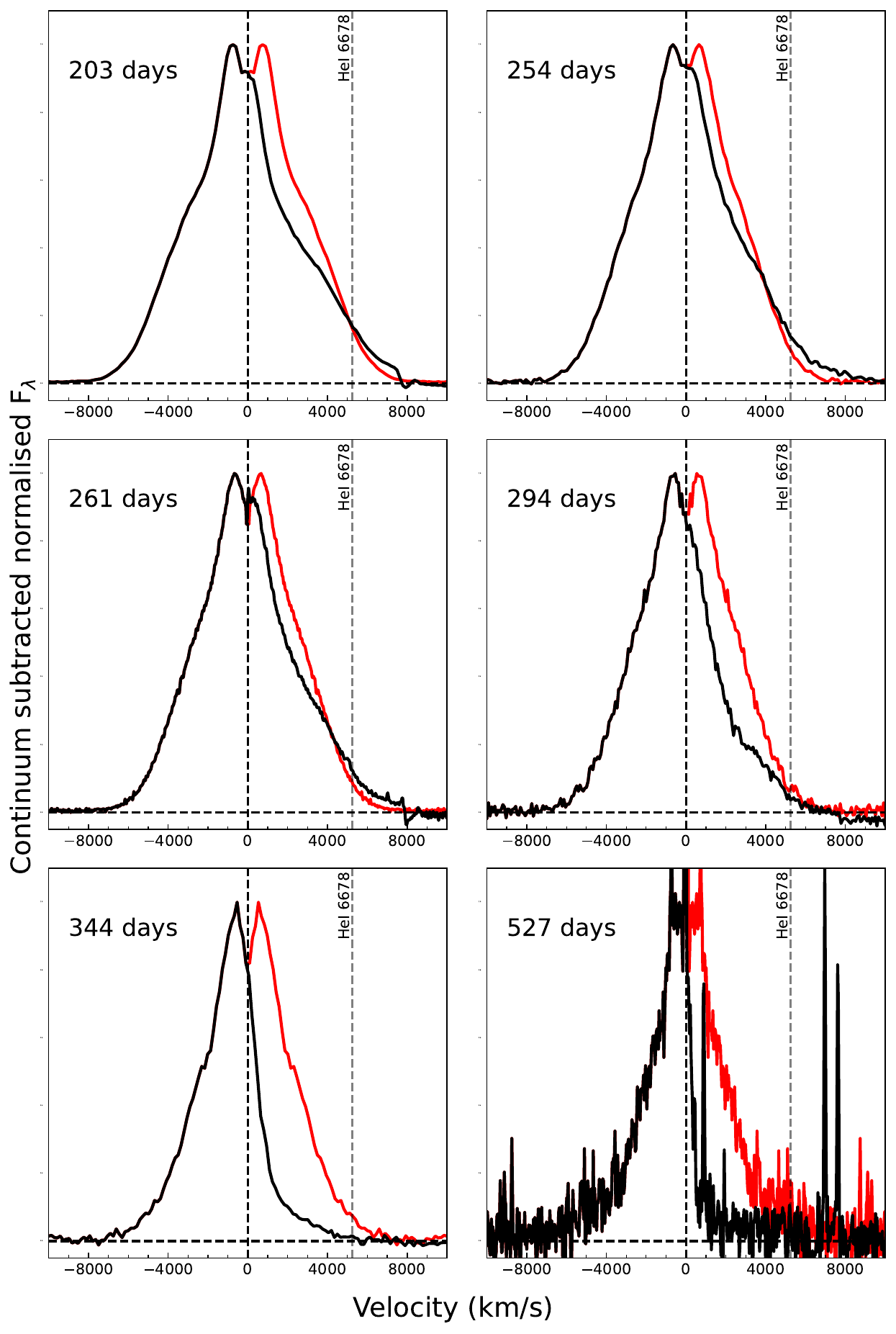}
      \caption{H$\alpha$ line profiles with the blue side of the emission line reflected around the rest wavelength of H$\alpha$. The continuum has been subtracted and the lines normalised to the peak of the broad H$\alpha$.
      }
\label{fig:Ha_reflection}
\end{figure}

The spectra of SN~2021irp evolve slowly, with the main features not changing drastically during the period between 200~d and 345~d. The most visible change is the development of a large asymmetry in the broad lines. Fig. \ref{fig:Halpha comp} shows this phenomenon for the H$\alpha$ line. There is a slight decrease in the line width on the blue side over the 200~d - 345~d period, followed by a further decrease at 528~d, potentially corresponding to a gradual decline in the velocity of the emitting gas. On the red side of the line, the shape of the line barely evolves between 200~d and 261~d, but at 294~d has dramatically decreased in strength compared to the blue wing, which continues to barely evolve in shape. The trend continues to our spectrum at 345~d and finally to our latest spectrum at 528~d, which still shows a broad component on the blue side of the emission feature, and shows further erosion of the emission on the red side. The line asymmetry is also shown in Fig. \ref{fig:Ha_reflection}, where we mirror the line around the rest wavelength of H$\alpha$. It is clear here that the line is not simply shifted, but actually asymmetric, particularly in the final two epochs. The line asymmetry is also evident in the H$\beta$, \ion{He}{I} $\lambda$5876~\AA, Ca NIR triplet, and \ion{Fe}{II} $\lambda$5018~\AA~emission features, although line blending and lower signal to noise makes it less clear. This phenomenon is consistent with dust formation and the timing of the change in shape of the red wing is consistent with the timing of the change in colour and decline rate observed in the photometry. We discuss this scenario in Sect. \ref{sec:dust_formation}.

The final spectrum taken at 528~d  still does not display the typical nebular spectrum common for Type II SNe observed at late times \citep[see e.g.][]{Jerkstrand2017}, and the commonly observed strong forbidden emission lines of [\ion{O}{I}] $\lambda\lambda$6300, 6364~\AA~ are not detected. To estimate the velocity of the fastest moving ionised material, we measure the blue edge at zero intensity of the H$\alpha$ emission line by fitting and subtracting the continuum 
and taking the first negative flux value in the continuum subtracted spectrum to represent the blue edge. This yields $\sim5000$~km~ s$^{-1}$, which corresponds to a distance of 2.3$\times10^{16}$ cm at this epoch. As mentioned above, the red wing is highly eroded and the red edge at zero intensity is at 1100~km~s$^{-1}$.

\begin{figure}
   \centering
   \includegraphics[width=0.5\textwidth]{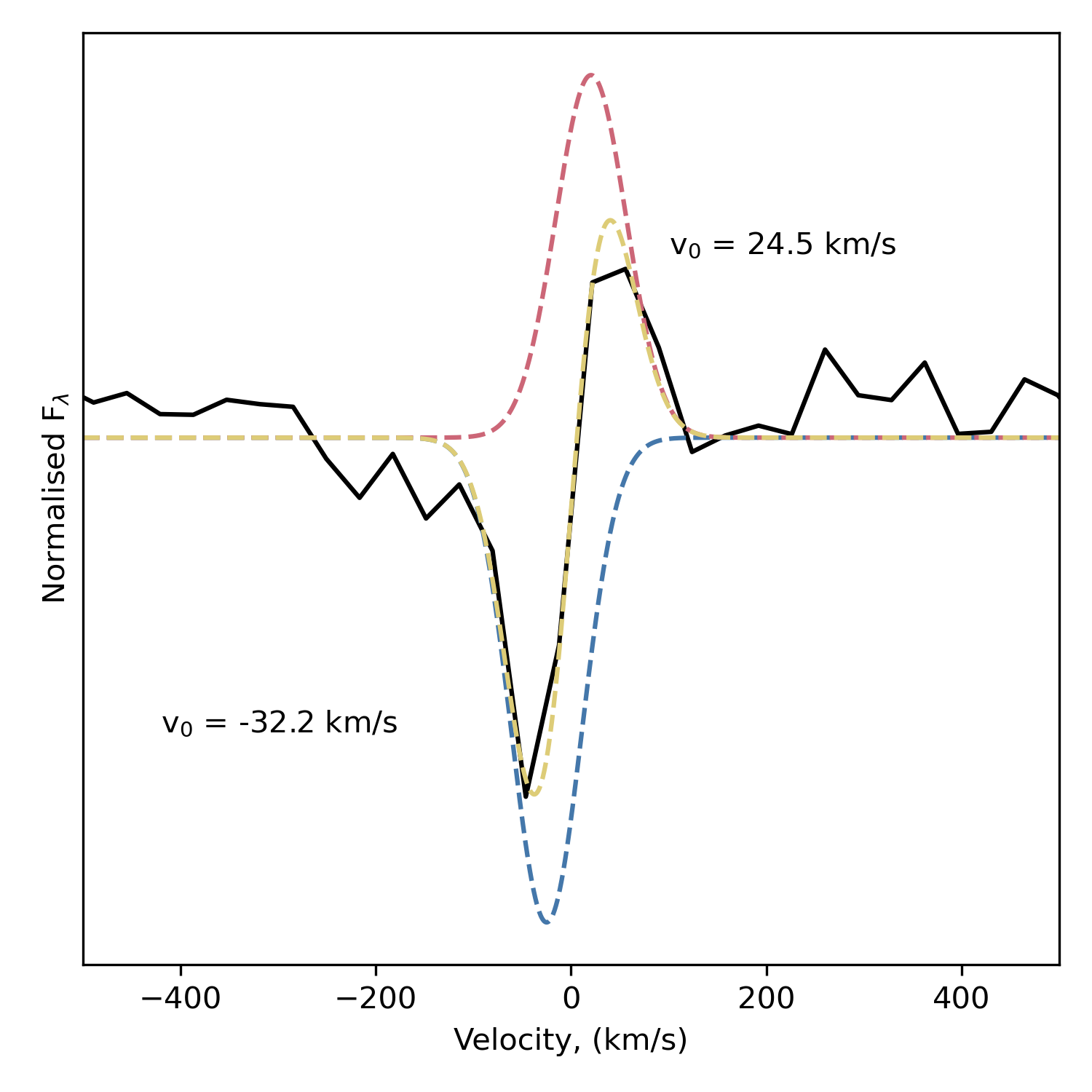}
      \caption{The apparent narrow P-Cygni profile in the spectrum taken at 262d. We subtracted the overlaying broad emission feature and fit two Gaussian profiles with FWHMs fixed to the instrumental resolution of 85~km~s$^{-1}$ to the residual spectrum. The red and blue dotted lines show the best-fit emission and absorption features, respectively, while the yellow dotted line shows the sum of the two features. The peak-to-peak separation of the two profiles is 56.7 km/s.}
\label{fig:narrow_Pcygni}
\end{figure}

In our highest resolution spectrum of the H$\alpha$ region, we observe a narrow P-Cygni profile at the peak of the emission feature, shown in Fig. \ref{fig:narrow_Pcygni}. This feature has been observed in high-resolution spectroscopy of many interacting SNe and is interpreted as arising in the unshocked CSM \citep[see e.g.][]{Smith2020}. In this case, the absorption measures the velocity of the CSM. Fitting both the emission and absorption features with Gaussian profiles fixed to the instrumental resolution of 85~km~s$^{-1}$, measured from the width of emission lines arising from a nearby \ion{H}{II} region, yields an offset between the two of $\sim1$~\AA ~, corresponding to a velocity of $\sim56.7 \pm 0.2$ km s$^{-1}$. This is smaller than the resolution of the instrument, so we can not confidently claim this as the velocity of the CSM and instead consider the velocity to be limited to less than the instrumental resolution of 85~km~s$^{-1}$.

\subsection{\label{subsec:spec_comparison} Spectral comparison}

\begin{figure*}
   \centering
   \includegraphics[width=0.49\textwidth]{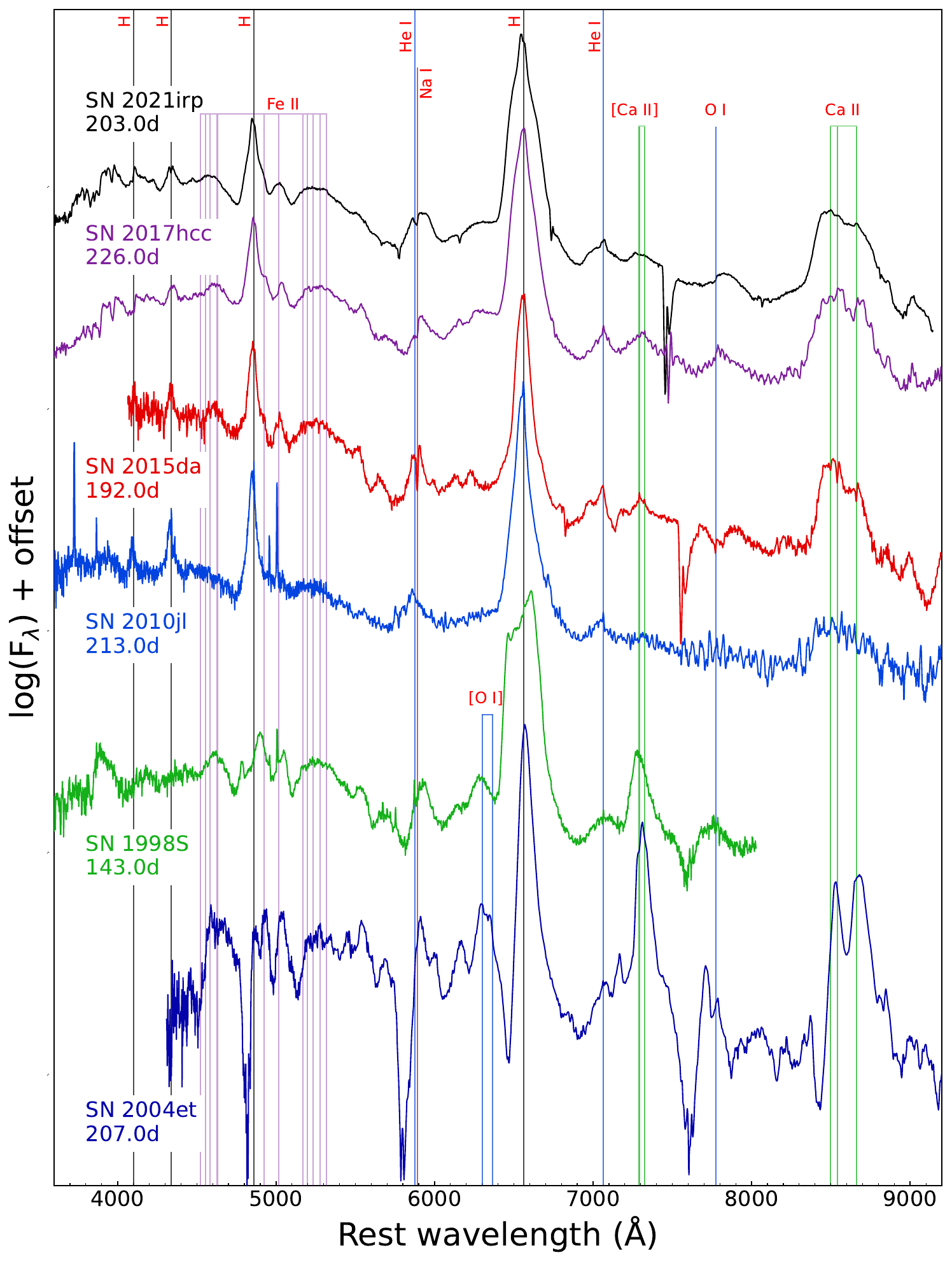}
   \includegraphics[width=0.49\textwidth]{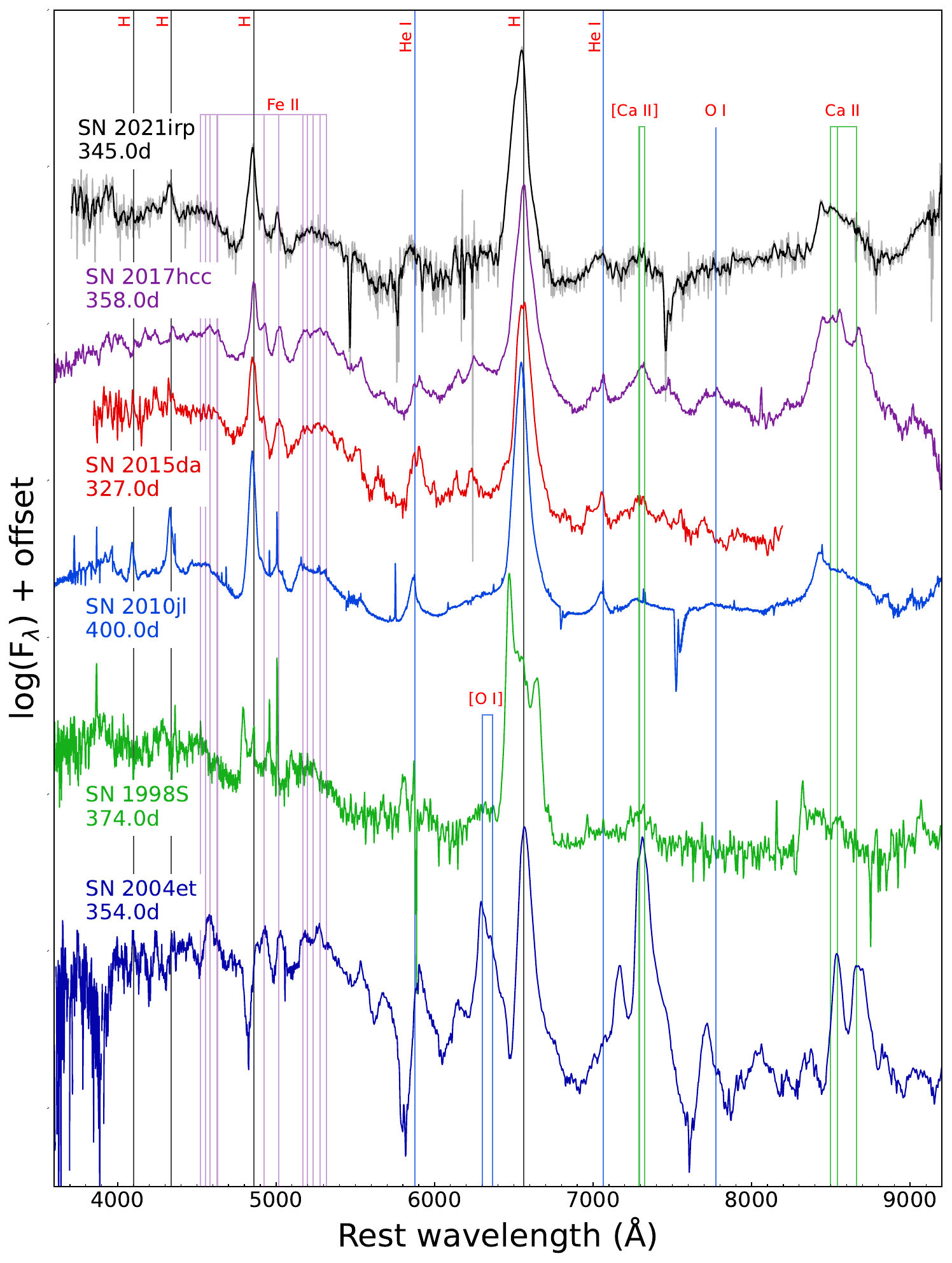}
      \caption{Spectra of SN 2021irp alongside selected SNe at the closest available epoch. Spectra have been shifted to aid visual comparison. The spectrum of SN~2021irp at 345~d has been slightly smoothed for visual clarity. \textbf{Left panel:} Spectra at $\sim200$~d. \textbf{Right panel:} Spectra at $\sim350$~d. }
\label{fig:spec_comp}
\end{figure*}

In Fig. \ref{fig:spec_comp}, we compare selected spectra of SN~2021irp to other selected SNe obtained from WISeREP \citep{Yaron2012}. The associated references and physical parameters are listed in Tab. \ref{tab:comparison_objects}. At approximately 203~d, SN~2021irp has a strong resemblance to the other interacting SNe in our sample (SN~2010jl, SN~2015da, and SN~2017hcc) at similar epochs, with the continuum shape, observed emission lines and their widths being strikingly similar. SN~2004et has a very different spectrum, as expected for a Type IIP supernovae. SN~1998S is transitional between the strongly interacting SNe and the weakly interacting SN~2004et, lacking absorption features and showing an \ion{Fe}{II} pseudo-continuum like the former, and having strong nebular lines of [\ion{O}{I}] $\lambda\lambda$6300,6364~\AA~and [\ion{Ca}{II}] $\lambda$7291,7323~\AA~like the latter. Later, at approximately 345~d, the spectra of the interacting SNe are still similar. SN~2021irp shows a rising continuum at wavelengths redder than 8000~\AA, due to having a strong IR component. This can be observed in the other interacting SNe, but is most pronounced in SN~2021irp and SN~2010jl.

\begin{figure}
   \centering
   \includegraphics[width=0.5\textwidth]{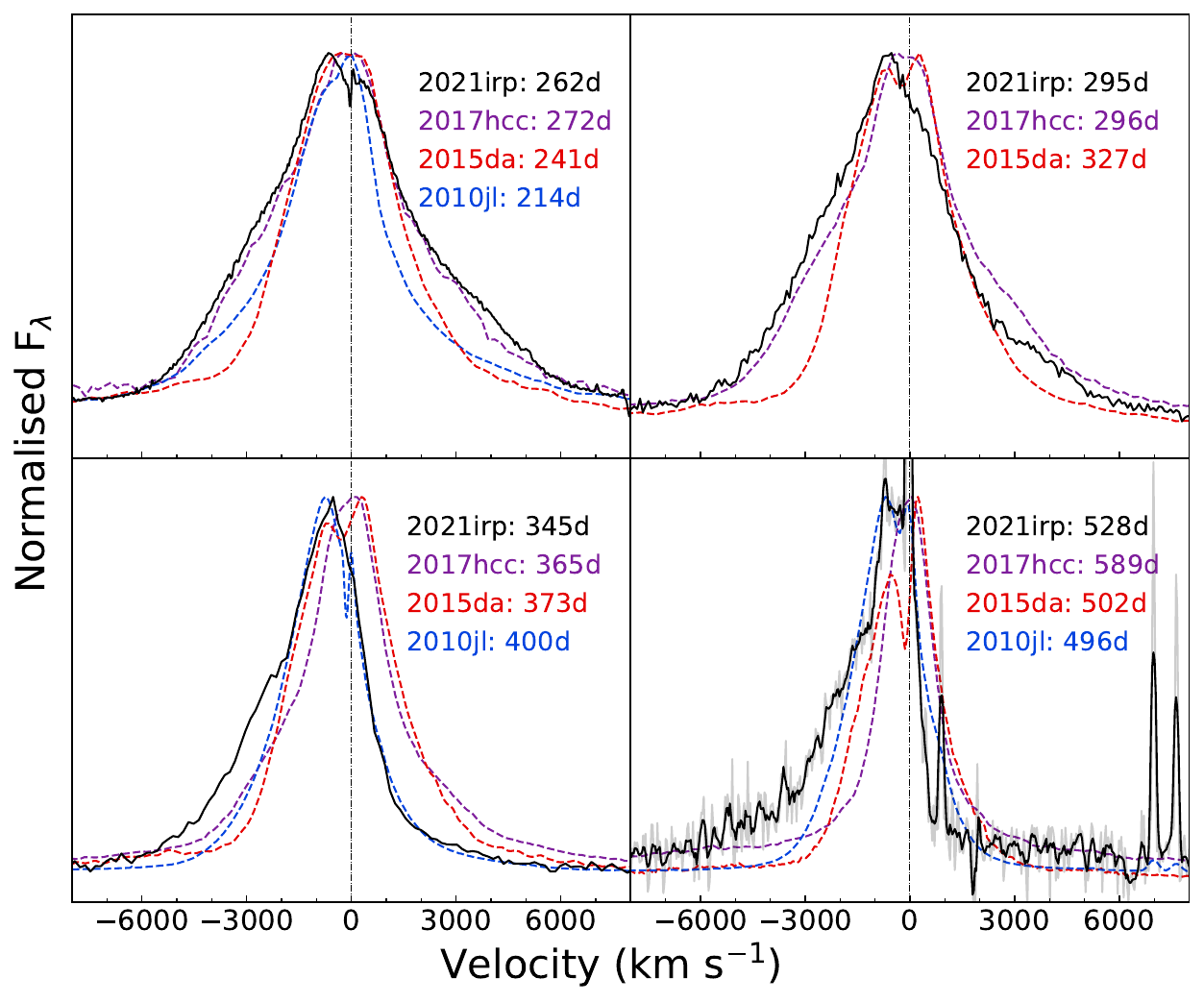}
      \caption{H$\alpha$ line profiles of SN~2021irp and other selected interacting SNe at similar epochs. The peaks of the emission line are normalised to one using the maximum flux in the H$\alpha$ region, except for in the 528 d spectrum of SN~2021irp, where we use the section between the narrow H$\alpha$ and narrow  [\ion{N}{II}] $\lambda$6548 lines which may arise from the galaxy. The latest spectrum of SN~2021irp is slightly smoothed for clarity.}
      
\label{fig:Ha_spec_comp_21irp}
\end{figure}

The interacting SNe display a variety of H$\alpha$ profiles, as shown in Fig. \ref{fig:Ha_spec_comp_21irp}. The line profiles of SN~2021irp and SN~2017hcc are very similar before the red side of the emission line begins to be eroded, after which SN~2017hcc does not display the erosion of the red side of the line to any clear extent. Also notably, the blue side of the emission line for SN~2017hcc becomes narrower over time, whereas it stays relatively constant for SN~2021irp. We note that, as commented on in \citet{Smith2020}, there is some asymmetry in the line profiles of SN~2017hcc, but it appears to be less than in SN~2021irp. The emission lines of SN~2010jl show strong asymmetry at 400~d, where the red side of the line profile is similar to that of SN~2021irp. The blue wing is similar at the peak of the profile (i.e. the intermediate component), but lacks the broad base of the line displayed by SN~2021irp.

\section{Discussion}
\label{sec:discussion}
\subsection{Energy source}\label{subsec:energy_source}

SN~2021irp shows spectra that have many features in common with photospheric-phase spectra of Type II SNe, with broad emission lines of the Balmer series, \ion{He}{I} $\lambda$5876~\AA~, \ion{Fe}{II} and the Ca NIR triplet. However, the duration of the bright photospheric phase is much longer ($>350$~d) than normal Type IIP SNe ($\sim 100$~d, \citep{Anderson2014}). Even at $\sim 350$~d after the explosion, it shows not a nebular but a photospheric spectrum with an absolute magnitude of $o\sim -16$ mag. At this late phase, a typical Type IIP SNe will be nebular and powered by the decay of $^{56}$Co. A typical absolute magnitude would be $\sim - 11\pm1$ mag, taking the mean observed magnitude at the beginning of the tail phase of $-13.66\pm0.83$ from \citet{Anderson2014} and assuming the expected tail decline rate of $\sim1$~mag per 100~d for $^{56}$Co decay assuming full emission trapping. Additionally, the total radiated energy in SN~2021irp, > 2.6$\times10^{50}$~erg, is much larger than the typical values for Type IIP SNe ($\sim 10^{49}$~erg for the plateau phase). Therefore, this SN must have at least one other energy source in addition to the typical energy sources for the radiation in Type IIP SNe i.e. the thermal energy deposited by the explosion shock and the radioactive energy from a typical amount ($\lesssim 0.1$ M$_{\odot}$) of $^{56}$Ni and $^{56}$Co.

Several mechanisms have been proposed in order to explain the huge amount of radiation in some very luminous SNe \citep{GalYam2019,Kangas2022,Kangas2024}, including energy input from the decay of several solar masses of $^{56}$Ni \citep{Kasen2011,Terreran2017}; from central engines such as a fast-spinning magnetar \cite{Kasen2010, Woosley2010}; fallback accretion on to a newly born black hole \citep{Dexter2013}; and interaction between the SN ejecta and the circumstellar material (CSM; \citealt{Chevalier1994}).

The scenario with a large amount of $^{56}$Ni, which might be predicted in a pair-instability SNe, is inconsistent with the light curve shape in SN~2021irp. The light curve shapes predicted in pair-instability SNe are, in general, composed of an early decline from shock cooling, a diffusion bump, and a linear decline following the $^{56}$Co decay \citep[e.g.,][]{Moriya2010, Kasen2011, Dessart2013}, although the detailed behaviours depend on the progenitor properties. We do not see this kind of light curve evolution in SN~2021irp, whose light curve consists of at least two power laws with different decline rates (0.0076 mag/day and 0.027 mag/day before and after $\sim 230$ days, respectively). Here, the earlier decline of SN~2021irp is slower than the one due to the Co decay, while the later decline is faster. The predicted spectral features in pair-instability SNe \citep[][]{Kasen2011,Jerkstrand2016} are also inconsistent with the observed spectra of SN~2021irp.

Central engine scenarios have also difficulties in explaining the photometric and spectroscopic properties of SN~2021irp. In particular, the light curve decline rate at later phases (e.g., 0.027 mag/day at $\sim 300$ days) does not match with the values predicted for magnetar power \citep[$t^{-2}$, e.g.,][]{Dessart2018} or for fallback accretion \citep[$t^{-5/3}$, e.g.,][]{Dexter2013}. In addition, the spectra dominated by emission lines in SN~2021irp are also difficult to reproduce with a central engine scenario \citep[e.g.,][]{Dessart2018}. If the ejecta is optically thick and heated from the deep inside as in the central engine scenarios, the ejecta would naturally have absorption features, especially in strong lines \citep[e.g.,][]{Dessart2018}.

Another important constraint on the energy source comes from the location of the emitting region of the ejecta. Since SN~2021irp shows photospheric spectra even at late phases (see Fig.~\ref{fig:spectra}), its ejecta are not yet completely optically thin and thus have a photosphere. From analogy with normal Type~II SNe, the location of the photosphere should be around the emitting regions of prominent broad lines (e.g., H$\alpha$, H$\beta$, Ca II triplet), which corresponds to $r = 4000$ km s$^{-1} \times 200$~(350) days $ = 7 \times 10^{15}$ ($1.2 \times 10^{16}$) cm at 200 (350) days after explosion. These values are much larger than the estimated blackbody radius (see Fig.~\ref{fig:BB_bolo}). This suggests that the photosphere is not spherical as in usual Type~II SNe and instead the emitting regions exist locally in the outer parts of the SN ejecta. We can estimate the ‘filling factor' defined as the ratio of the surface area of the emitting region derived from blackbody fitting and the emitting region implied from the photosphere location derived from the line widths. 
At 200 d, our measured blackbody radius is $1.7 \times 10^{15}$ cm and thus the filling factor is $(1.7 \times 10^{15}/ ~7 \times 10^{15})^2 = 0.059$ => $\sim 6$\%. Since the radiation produced in the outer layers can avoid forming large absorption features, this can additionally explain why the broad lines have emission-dominated shapes rather than P-Cygni shapes. In the context of CSM interaction, this picture of a patchy photosphere can be achieved if the SN ejecta interacts with an aspherical CSM. Furthermore, this can also explain why SN~2021irp does not have narrow Balmer lines arising from the ionised CSM, which is regarded as a common and important signature of CSM interaction. The locally-ionised regions of ejecta around the aspherical CSM interaction create local optically-thick photospheres, which might hide the emitting regions of the narrow Balmer lines from the ionised-unshocked CSM.

Therefore, we conclude that the main energy source for SN~2021irp is CSM interaction. We quantitatively discuss the exact CSM parameters to explain the observational properties in Paper~II. 

\subsection{Dust formation}\label{sec:dust_formation}

Core-collapse SNe are thought to be important producers of dust, which is a basic ingredient for planets, stars, and galaxies in the universe. Dust is theoretically predicted to be formed in inner SN ejecta \citep[e.g.,][]{Hoyle1970, Nozawa2003, Dwek2007, Sarangi2015} and also in the cool dense shell of interacting SNe \citep[e.g.,][]{Pozzo2004,Mattila2008,Sarangi2018}. This idea is supported by IR excesses, asymmetries in emission-line profiles, and / or acceleration of the light curve declines observed in Type II SNe \citep[e.g.,][]{Lucy1989,Danziger1989,Pozzo2004, Sugerman2006,Kotak2009, Andrews2010, Andrews2011a, Fabbri2011, Meikle2011, Szalai2011,Matsuura2011, Tinyanont2019, Bevan2019, Shahbandeh2023}. The dust masses estimated at several hundred days after explosion in Type II SNe \citep[see][and references therein]{Tinyanont2019, Shahbandeh2023} and in interacting SNe \citep[e.g.,][]{Mattila2008,Stritzinger2012, Maeda2013, Gall2014, Uno2023, Wang2024,Shahbandeh2024} are $10^{-5}$ - $10^{-2}$ M$_{\odot}$ yr$^{-1}$.

A dramatic feature in the light curves of SN~2021irp is the acceleration of the decline at $\sim 230$ days (see Fig.~\ref{fig:LC}). Since the main energy source for the radiation is the CSM interaction, this decline could be interpreted as a rapid decrease of the energy input, i.e., the steepening of the CSM distribution. However, another possibility for the change of the decline rate could be the formation of dust. There are a number of observational features of SN~2021irp that point to this interpretation: 

\begin{itemize}
    \item[1] The increase in the decline of the SN luminosity, as shown in Fig. \ref{fig:bolo_comp},
    \item[2] The change in the SN colour, with particularly the $V-i$ colour sharply turning to the red, as shown in Fig. \ref{fig:colors},
    \item[3] The erosion of the red wing of the emission lines, particularly clearly seen in H$\alpha$,
    \item[4] The luminous and long-lasting IR excess.
\end{itemize}

Crucially, all four of these features first occur (or reach their peak luminosity, in the case of the IR excess) at the same time, between 250 and 300d, implying that they could all be caused by the same physical phenomenon. Dust formation can explain all of these observations \citep[see e.g.][]{Smith2012}, and we discuss them in turn.

Freshly formed dust can absorb the radiation from the SN, which will lead to a decrease in the observed luminosity in the optical. The dust will then re-radiate this energy in the IR. In Fig. \ref{fig:BB_bolo}, it is clear that as the luminosity arising from the optical blackbody (i.e. the heated SN ejecta) decreases, the IR luminosity arising from the dust increases, such that the total luminosity evolves smoothly. This implies that there is not necessarily a need for a rapid decrease in the energy input. Absorption by dust has a strong wavelength dependence, with larger absorption at bluer wavelengths, which contributes to the colour change towards the red that we observe. The colour change is additionally driven by emission from the dust in the $i$ band, which is visible in Fig. \ref{fig:SED_fitting_IR}.

The erosion of the red wing of emission lines, or a blueshift in the emission line profile, is commonly seen in Type IIn SNe and has a number of proposed explanations \citep[see e.g.][]{Smith2020}. Dust formation can explain this feature, as newly formed dust preferentially obscures the receding SN ejecta as opposed to the oncoming ejecta, eroding the red wing of the emission line. In SN~2010jl, \citet{Fransson2014} observed symmetric Lorentzian profiles with a systematic blueshift, and therefore argue that the blueshift is caused by radiatively accelerated CSM. This cannot explain the observations for SN~2021irp as the emission lines are not symmetric Lorentzians with any shift of centroid - see the reflected profile at 351d in Fig. \ref{fig:Ha_reflection}. Alternatively, SN~2010jl has been interpreted as a dust-forming SN \citep{Gall2014,Sarangi2018}. Another proposed explanation is that, instead of dust, the continuum photosphere can obscure the receding ejecta. However, this would suggest that the effect becomes weaker over time (as the photosphere becomes smaller), and we observe the opposite in SN~2021irp. Another prediction of the dust formation scenario is that there should be a wavelength dependence in the line erosion, with bluer lines experiencing a stronger effect, although the line erosion depends on the locations of the emitting regions and the dust. We do in fact see some increase in the Balmer decrement (see Fig. \ref{fig:Balmer_measurements}), which could correspond to this effect.

As described in Sect. \ref{sec:BB_fits}, the IR excess we observe has properties consistent with emission from dust. The dust could be newly formed, but there is another possible origin: an IR echo due to pre-existing circumstellar dust \citep[e.g.,][]{Bode1980,Graham1983,Dwek1983,Meikle2006,Meikle2007,Maeda2015}. In particular, in the case of SN~2021irp there is a large amount of CSM which could also contain circumstellar dust, as is observed in the material around massive stars \citep[see e.g][]{Gehrz1971,Smith2014}. In the case where the IR excess arises from pre-existing dust, we expect to observe an IR echo due to the heating of the dust by the radiation from the SN as has been previously observed in a number of interacting SNe \citep[e.g., ][]{Tartaglia2020,Dwek2021,Moran2023}. The estimated temperature of the dust is $\sim 1700$ K at 274 days, which requires carbon dust rather than silicate to avoid sublimation, and the dust contributing to the echo should be located around $\sim 0.02$ pc from the SN, assuming radiative equilibrium of the absorption and re-emission by the dust grains, and our measured peak luminosity of SN~2021irp \citep[see Equation (4) in][]{Maeda2015}. This distance corresponds to a light crossing time of only $\sim 20$ days. This is not consistent with the observed timing and duration of the IR excess, which lasts many 100s of days.
Furthermore, the correspondence of the emergence of the IR excess with the change of the light curve decline rate and with the erosion of the H$\alpha$ line supports the argument that the main contributor for the IR excess is newly-formed dust.

An IR echo might significantly contribute to the IR excess at early phases, as in the case of the Type~IIn SN 2010jl \citep[][]{Andrews2011b,Fransson2014,Bevan2020,Dwek2021}. The IR excess at 161~d shows a lower blackbody temperature and larger BB radius than those for the later epochs (see Figure~\ref{fig:BB_bolo}). This lower temperature can be explained with an IR echo from more distant pre-existing dust rather than newly formed hot dust, which is inferred for the IR excesses at later times. In addition, the large blackbody radius cannot be explained only with the newly formed dust. The blackbody radius indicates the lower limit for the distance of the emitting dust from the SN. Since even the outermost layer of the ejecta with 10000 km~s$^{-1}$ can only expand to $1.4 \times 10^{16}$ cm away from the centre of the ejecta at 161~d, and the blackbody radius is $2.4 \times 10^{16}$ cm at that epoch, the newly formed dust, which is created in the inner ejecta and/or the interaction shock shell, cannot create such a large emitting region as the radius observed at 161~d. On the other hand, our measurements of the dust at epochs later than 269~d have blackbody radii $>1.6\times~10^{16}$~cm, implying that SN ejecta with velocities $>6900$~km~s$^{-1}$ can reach the location of this dust. This is consistent with dust formation in the interaction shock shell for these epochs, while the first epoch is most likely dominated by emission from an IR echo.

Since most of the red wing of the H$\alpha$ line suffers from dust absorption, the dust responsible for the erosion of the line should be relatively spread out in space. This naturally occurs if the dust forms in the inner SN ejecta. Dust formation in the inner SN ejecta is inhibited at early times first by the high temperatures during the photospheric phase, and at later phases by the presence of hard radiation and non-thermal electrons arising from CSM interaction \citep{Sarangi2018}. For SN~1987A, observations showed that the dust formation began later than 250~d \citep{Bouchet1993,Wooden1993,Dwek2015}. The onset of dust formation in SN~2021irp occurs at 250~d-300~d, consistent with that observed for SN~1987A, implying that dust formation in the inner SN ejecta is possible at these epochs. However, the ongoing CSM interaction is much stronger for SN~2021irp, which might prevent dust formation in the ejecta.

The large blackbody radius observed for the IR excess at >269~d, implies that only SN ejecta at relatively high velocities (>6900 km~s$^{-1}$ at 269~d) have reached the location of the emitting dust. Dust can not form in the SN ejecta at these velocities, as there are not sufficient refractory elements in this part of the ejecta \citep[see e.g.][]{Mattila2008,Sarangi2018}. This implies that the dust must be forming in the cool dense shell in the interaction region \citep{Pozzo2004,Mattila2008}. It is not possible to ascertain if this dust is the dominant cause of the erosion in the emission lines, and therefore that dust formation in the SN ejecta is not required, without further constraining the geometry of the asymmetric CSM and line-forming regions.

In summary, to explain our observations of SN~2021irp, we require newly formed dust in the interaction shock shell to explain the observed blackbody radius; there could be dust in the SN ejecta, which would most easily explain the observed erosion of the red part of the line; and an IR echo is the most likely explanation to explain the first IR epoch.

In order to estimate the total amount of the newly formed dust ($M_{\rm{dust}}$) at $\sim 700$~d, we assume single-size ($a=0.1$ $\mu$m) graphite grains with a single temperature ($T_{\rm{dust}} \sim 1000$ K). The total luminosity from the dust grains is expressed in the optically thin limit:
\begin{equation}
    L_{\rm{dust}} = 4 \pi M_{\rm{dust}} \int \kappa_{\rm{abs},\nu} B_{\nu}(T_{\rm{dust}}) \rm{d}\nu = 4 \sigma_{\rm{SB}} \overline{\kappa}_{\rm{abs}} M_{\rm{dust}} T_{\rm{dust}}^{4},
\end{equation}
where $\sigma_{\rm{SB}}$ is the Stefan–Boltzmann constant. Here we, for simplicity, ignore the frequency dependence of the opacity of the dust and use the constant value of the opacity at the frequency of the peak of the Planck function at this dust temperature ($\overline{\kappa}_{\rm{abs}}$). The peak frequency of the Planck function is $\sim 6 \times 10^{13}$ Hz for $T_{\rm{dust}} \sim 1000$ K, and thus $\overline{\kappa}_{\rm{abs}} \sim 10^{3}$ cm$^2$~g$^{-1}$ for graphite grains with 0.1 $\mu$m and with $T_{\rm{dust}} \sim 1000$ K. We also assume that the total observed luminosity at $\sim 700$ days ($\sim 3 \times 10^{41}$ erg; see Fig.~\ref{fig:BB_bolo}) is dominated by the dust emission and thus it represents the total luminosity from the dust grains. Under the above assumptions, we estimate the dust mass as $M_{\rm{dust}}\sim 10^{-3}$ M$_{\odot}$. This value is consistent with the masses of the newly formed dust in Type~IIP SNe and interacting SNe at similar epochs\citep[see e.g.][and refs above]{Shahbandeh2023,Sarangi2018,Kotak2009}. We note that, if the dust is formed in optically thick regions, the actual amount should be higher than the derived value here.

\section{Summary and conclusions}\label{sec:conclusions}

In this work, we have presented results from our spectroscopic and photometric follow-up observations of the luminous Type II SN~2021irp. We summarise our findings as follows: 

\begin{enumerate}
  \item Despite missing much of the early evolution, we determine that SN~2021irp is a very luminous SN based on its brightness at 200~d and later, and find it to have a radiated energy too large to explain for a typical Type II SN powered by the release of thermal energy from the ejecta and radioactive decay from synthesised $^{56}$Ni/$^{56}$Co without an additional energy source.
  \item We consider the various sources of additional energy and find that CSM interaction is the natural candidate, based on the photometric and spectroscopic properties of SN~2021irp. In particular, the pseudo-bolometric luminosity evolution and non-spherical photosphere shape that we infer from the broad spectral lines are inconsistent with other possible scenarios.
  \item We determine that the CSM powering the SN must exhibit significant asymmetry, to produce the observed broad emission line features and the observed evolution of the BB radius. Furthermore, we observe the presence of unshocked CSM and constrain the velocity of the pre-SN mass loss to $\lesssim$ 85 km~s$^{-1}$.
  \item We find extensive evidence of the formation of dust, and can constrain the time when the formation begins to $\sim 250$~d. We estimate the dust mass formed as $\sim10^{-3}$~M$_{\odot}$, consistent with masses of newly formed dust observed in Type IIP and interacting SNe. The dust is required to form in the interaction shock shell and some dust may also form in the SN ejecta. We find possible evidence of an IR echo at phases earlier than the onset of dust formation, arising from pre-existing dust that exists in the unshocked CSM.
\end{enumerate}

We continue our analysis of SN~2021irp in Paper II, presenting imaging- and spectro-polarimetry and conducting light curve modelling. Using these results along with those presented here, we are able to quantify the properties of the CSM that powers the SN.

\begin{acknowledgements}
T.M.R is part of the Cosmic Dawn Center (DAWN), which is funded by the Danish National Research Foundation under grant DNRF140. T.M.R acknowledges financial support from the Finnish Academy of Science and Letters through the Finnish postdoc pool. T.M.R and S. Mattila acknowledge support from the Research Council of Finland project 350458. 
We are grateful to Jan Aaltonen, Ville Antila, Katja Matilainen and Sofia Suutarinen, who observed this target as part of the "NOT course 2021" organised by the Department of Physics and Astronomy at the University of Turku in October 2021.
T.N. thanks Masaomi Tanaka and Akihiro Suzuki for fruitful discussion. T.N. and H.K. acknowledge support from the Research Council of Finland projects 324504, 328898, and 353019.
S. Moran acknowledges support from the Magnus Ehrnrooth Foundation and the Vilho, Yrj\"{o} and Kalle V\"{a}is\"{a}l\"{a} Foundation. 
R.K. acknowledges support from the Research Council of Finland (grant 340613).
T.K. acknowledges support from the Research Council of Finland project 360274.
C.P.G. acknowledges financial support from the Secretary of Universities and Research (Government of Catalonia) and by the Horizon 2020 Research and Innovation Programme of the European Union under the Marie Sk\l{}odowska-Curie and the Beatriu de Pin\'os 2021 BP 00168 programme, the support from the Spanish Ministerio de Ciencia e Innovaci\'on (MCIN) and the Agencia Estatal de Investigaci\'on (AEI) 10.13039/501100011033 under the PID2023-151307NB-I00 SNNEXT project, from Centro Superior de Investigaciones Cient\'ificas (CSIC) under the PIE project 20215AT016 and the program Unidad de Excelencia Mar\'ia de Maeztu CEX2020-001058-M, and from the Departament de Recerca i Universitats de la Generalitat de Catalunya through the 2021-SGR-01270 grant. 
Y.-Z. Cai is supported by the National Natural Science Foundation of China (NSFC, Grant No. 12303054), the National Key Research and Development Program of China (Grant No. 2024YFA1611603), the Yunnan Fundamental Research Projects (Grant No. 202401AU070063), and the International Centre of Supernovae, Yunnan Key Laboratory (No. 202302AN360001). 
K.M. acknowledges support from the Japan Society for the Promotion of Science (JSPS) KAKENHI grant JP24KK0070 and 24H01810. The work is partly supported by the JSPS Open Partnership Bilateral Joint Research Projects between Japan and Finland (K.M. and H.K.; JPJSBP120229923)
A.R. acknowledges financial support from the GRAWITA Large Program Grant (PI P. D’Avanzo) and the PRIN-INAF 2022 "Shedding light on the nature of gap transients: from the observations to the models".
M.F. is supported by a Royal Society - Science Foundation Ireland
University Research Fellowship.
N.E.R. acknowledges support from the PRIN-INAF 2022, `Shedding light on the nature of gap transients: from the observations to the models
A.R. acknowledges financial support from the GRAWITA Large Program Grant (PI P. D’Avanzo) and the PRIN-INAF 2022 "Shedding light on the nature of gap transients: from the observations to the models".
This work was partly supported by Grant-in-Aid for Scientific Research (C) 22K03676.
We acknowledge ESA Gaia, DPAC and the Photometric Science Alerts Team\footnote{\url{http://gsaweb.ast.cam.ac.uk/alerts}}.
This publication makes use of data products from the Two Micron All Sky Survey, which is a joint project of the University of Massachusetts and the Infrared Processing and Analysis Center/California Institute of Technology, funded by the National Aeronautics and Space Administration and the National Science Foundation.
The spectrum taken by the Seimei telescope was obtained under the KASTOR (Kanata And Seimei Transient Observation Regime) project, specifically under the following program for the Seimei Telescope at the Okayama observatory (21B-O-0009). The Seimei telescope at the Okayama Observatory is jointly operated by Kyoto University and the National Astronomical Observatory of Japan (NAOJ), with assistance provided by the Optical and Infrared Synergetic Telescopes for Education and Research (OISTER) program.
The Pan-STARRS1 Surveys (PS1) and the PS1 public science archive have been made possible through contributions by the Institute for Astronomy, the University of Hawaii, the Pan-STARRS Project Office, the Max-Planck Society and its participating institutes, the Max Planck Institute for Astronomy, Heidelberg and the Max Planck Institute for Extraterrestrial Physics, Garching, The Johns Hopkins University, Durham University, the University of Edinburgh, the Queen's University Belfast, the Harvard-Smithsonian Center for Astrophysics, the Las Cumbres Observatory Global Telescope Network Incorporated, the National Central University of Taiwan, the Space Telescope Science Institute, the National Aeronautics and Space Administration under Grant No. NNX08AR22G issued through the Planetary Science Division of the NASA Science Mission Directorate, the National Science Foundation Grant No. AST-1238877, the University of Maryland, Eotvos Lorand University (ELTE), the Los Alamos National Laboratory, and the Gordon and Betty Moore Foundation.
Funding for the Sloan Digital Sky Survey IV has been provided by the Alfred P. Sloan Foundation, the U.S. Department of Energy Office of Science, and the Participating Institutions. SDSS-IV acknowledges support and resources from the Center for High Performance Computing  at the University of Utah. The SDSS website is www.sdss4.org. SDSS-IV is managed by the Astrophysical Research Consortium for the Participating Institutions of the SDSS Collaboration including the Brazilian Participation Group, the Carnegie Institution for Science, Carnegie Mellon University, Center for Astrophysics | Harvard \& Smithsonian, the Chilean Participation Group, the French Participation Group, Instituto de Astrof\'isica de Canarias, The Johns Hopkins University, Kavli Institute for the Physics and Mathematics of the Universe (IPMU) / University of Tokyo, the Korean Participation Group, Lawrence Berkeley National Laboratory, Leibniz Institut f\"ur Astrophysik Potsdam (AIP),  Max-Planck-Institut f\"ur Astronomie (MPIA Heidelberg), Max-Planck-Institut f\"ur Astrophysik (MPA Garching), Max-Planck-Institut f\"ur Extraterrestrische Physik (MPE), National Astronomical Observatories of China, New Mexico State University, New York University, University of Notre Dame, Observat\'ario Nacional / MCTI, The Ohio State University, Pennsylvania State University, Shanghai Astronomical Observatory, United Kingdom Participation Group, Universidad Nacional Aut\'onoma de M\'exico, University of Arizona, University of Colorado Boulder, University of Oxford, University of Portsmouth, University of Utah, University of Virginia, University of Washington, University of Wisconsin, Vanderbilt University, and Yale University.
This research has made use of the NASA/IPAC Infrared Science Archive, which is funded by the National Aeronautics and Space Administration and operated by the California Institute of Technology. 
This research was partly based on observations made with the Nordic Optical Telescope (program IDs: P64-507 \& 66-506) owned in collaboration by the University of Turku and Aarhus University, and operated jointly by Aarhus University, the University of Turku and the University of Oslo, representing Denmark, Finland and Norway, the University of Iceland and Stockholm University at the Observatorio del Roque de los Muchachos, La Palma, Spain, of the Instituto de Astrofisica de Canarias.
The data presented here were obtained in part with ALFOSC, which is provided by the Instituto de Astrofisica de Andalucia (IAA) under a joint agreement with the University of Copenhagen and NOT.
This research was partly based on observations collected at the European Southern Observatory under ESO programmes 105.20DF.002, 108.228K.001 and 108.228K.002.
This publication makes use of data products from the Near-Earth Object Wide-field Infrared Survey Explorer (NEOWISE), which is a joint project of the Jet Propulsion Laboratory/California Institute of Technology and the University of California, Los Angeles. NEOWISE is funded by the National Aeronautics and Space Administration.
This work made use of Astropy: a community-developed core Python package and an ecosystem of tools and resources for astronomy \citep[Astropy Collaboration][]{astropy:2013,astropy:2018,astropy:2022}.
\end{acknowledgements}

%

\bibliographystyle{aa} 
\bibliography{bibliography} 

\begin{appendix}
\onecolumn
\section{Tables}
\begin{table}[H]
\caption{Photometry of SN~2021irp.  Photometry in ATLAS $co$ and SDSS $riz$ bands is presented in the AB system, while the remaining photometry is presented in the Vega system. Values are not corrected for extinction. Phases are in rest frame days with respect to our estimated explosion epoch. The full table is available at the CDS.}
    \centering
    \begin{tabular}{c c c c c c} \hline\hline
phase & mjd & mag & error & band & telescope+instrument \\ \hline
$-$22.59 & 59287.27 & >20.21 & - & c & ATLAS \\
$-$4.93 & 59305.28 & >19.92 & - & o & ATLAS \\
$-$2.97 & 59307.28 & >19.84 & - & o & ATLAS \\
$-$0.22 & 59310.08 & >20.7 & - & G & Gaia \\
2.90 & 59313.25 & 18.03 & 0.02 & c & ATLAS \\
18.58 & 59329.24 & 16.40 & 0.03 & o & ATLAS \\
20.54 & 59331.24 & 16.26 & 0.03 & o & ATLAS \\
109.19 & 59421.61 & 16.56 & 0.03 & o & ATLAS \\
132.71 & 59445.60 & 16.71 & 0.02 & o & ATLAS \\
132.73 & 59445.62 & 16.67 & 0.02 & o & ATLAS \\
134.69 & 59447.62 & 16.70 & 0.02 & o & ATLAS \\
138.13 & 59451.13 & 16.97 & 0.02 & G & Gaia \\
138.61 & 59451.62 & 16.71 & 0.01 & o & ATLAS \\
140.55 & 59453.59 & 16.76 & 0.01 & o & ATLAS \\
140.57 & 59453.61 & 16.73 & 0.04 & o & ATLAS \\
142.50 & 59455.58 & 16.75 & 0.03 & o & ATLAS \\
148.40 & 59461.60 & 16.76 & 0.02 & o & ATLAS \\
148.43 & 59461.63 & 16.79 & 0.01 & o & ATLAS \\
152.35 & 59465.62 & 16.82 & 0.00 & o & ATLAS \\
154.26 & 59467.57 & 17.54 & 0.01 & c & ATLAS \\
156.24 & 59469.58 & 16.90 & 0.01 & o & ATLAS \\
157.74 & 59471.11 & 13.86 & 0.04 & W1 & WISE \\
157.74 & 59471.11 & 13.16 & 0.06 & W2 & WISE \\
162.17 & 59475.63 & 16.87 & 0.02 & o & ATLAS \\
165.02 & 59478.53 & 16.86 & 0.02 & o & ATLAS \\
166.00 & 59479.54 & 16.90 & 0.02 & o & ATLAS \\
166.25 & 59479.80 & 17.17 & 0.02 & G & Gaia \\
168.00 & 59481.58 & 16.94 & 0.02 & o & ATLAS \\
173.87 & 59487.56 & 16.94 & 0.02 & o & ATLAS \\
173.95 & 59487.64 & 16.95 & 0.02 & o & ATLAS \\
177.80 & 59491.56 & 16.98 & 0.02 & o & ATLAS \\
179.73 & 59493.53 & 17.06 & 0.01 & o & ATLAS \\
183.74 & 59497.62 & 17.06 & 0.02 & o & ATLAS \\
187.56 & 59501.51 & 17.06 & 0.02 & o & ATLAS \\
191.48 & 59505.51 & 17.15 & 0.03 & o & ATLAS \\
193.43 & 59507.51 & 17.05 & 0.05 & o & ATLAS \\
199.41 & 59513.60 & 16.96 & 0.05 & o & ATLAS \\
199.80 & 59514.00 & 17.86 & 0.05 & V & NOT+ALFOSC \\
199.80 & 59514.00 & 17.35 & 0.04 & i & NOT+ALFOSC \\
199.80 & 59514.00 & 17.17 & 0.03 & r & NOT+ALFOSC \\
199.80 & 59514.00 & 16.90 & 0.08 & z & NOT+ALFOSC \\
199.80 & 59514.00 & 18.75 & 0.05 & B & NOT+ALFOSC \\
199.80 & 59514.00 & 16.99 & 0.03 & o & ATLAS \\
201.34 & 59515.57 & 17.15 & 0.02 & o & ATLAS \\
203.32 & 59517.58 & 17.11 & 0.02 & o & ATLAS \\
203.34 & 59517.61 & 17.11 & 0.05 & o & ATLAS \\
205.22 & 59519.52 & 18.00 & 0.03 & c & ATLAS \\
207.23 & 59521.57 & 17.17 & 0.03 & o & ATLAS \\
209.18 & 59523.56 & 17.97 & 0.04 & c & ATLAS \\
211.03 & 59525.45 & 17.26 & 0.02 & o & ATLAS \\
212.99 & 59527.44 & 18.12 & 0.02 & c & ATLAS \\
213.79 & 59528.25 & 17.49 & 0.03 & i & NOT+ALFOSC \\
213.79 & 59528.25 & 17.99 & 0.07 & V & NOT+ALFOSC \\
213.79 & 59528.25 & 18.87 & 0.04 & B & NOT+ALFOSC \\
213.79 & 59528.25 & 17.26 & 0.05 & r & NOT+ALFOSC \\
215.05 & 59529.55 & 17.28 & 0.03 & o & ATLAS \\
226.86 & 59541.58 & 17.41 & 0.08 & o & ATLAS \\
228.74 & 59543.50 & 17.38 & 0.02 & o & ATLAS \\
230.72 & 59545.52 & 17.45 & 0.04 & o & ATLAS \\
242.38 & 59557.40 & 17.61 & 0.02 & o & ATLAS \\
246.34 & 59561.45 & 17.65 & 0.03 & o & ATLAS \\
    \end{tabular}
    \label{tab:photometry_table}
\end{table}

\begin{table}[H]
\ContinuedFloat
\caption{continued}
    \centering
    \begin{tabular}{c c c c c c} \hline\hline
phase & mjd & mag & error & band & telescope+instrument \\ \hline
246.76 & 59561.87 & 17.91 & 0.06 & i & NOT+ALFOSC \\
246.76 & 59561.87 & 18.41 & 0.07 & V & NOT+ALFOSC \\
246.76 & 59561.87 & 19.26 & 0.04 & B & NOT+ALFOSC \\
246.76 & 59561.87 & 17.61 & 0.04 & r & NOT+ALFOSC \\
254.27 & 59569.53 & 17.49 & 0.13 & o & ATLAS \\
256.25 & 59571.54 & 17.80 & 0.01 & o & ATLAS \\
258.07 & 59573.40 & 17.90 & 0.02 & o & ATLAS \\
265.48 & 59580.95 & 18.32 & 0.04 & i & NOT+ALFOSC \\
265.48 & 59580.95 & 18.07 & 0.02 & o & ATLAS \\
265.48 & 59580.95 & 19.72 & 0.04 & B & NOT+ALFOSC \\
265.48 & 59580.95 & 18.90 & 0.04 & V & NOT+ALFOSC \\
265.48 & 59580.95 & 18.04 & 0.05 & r & NOT+ALFOSC \\
268.48 & 59584.02 & 15.39 & 0.12 & H & NOT+NOTCam \\
268.48 & 59584.02 & 16.33 & 0.09 & J & NOT+NOTCam \\
268.48 & 59584.02 & 14.54 & 0.15 & K & NOT+NOTCam \\
269.35 & 59584.90 & 18.17 & 0.06 & o & ATLAS \\
271.80 & 59587.40 & 19.02 & 0.09 & c & ATLAS \\
273.65 & 59589.29 & 18.31 & 0.05 & o & ATLAS \\
274.28 & 59589.93 & 18.13 & 0.22 & o & ATLAS \\
275.61 & 59591.28 & 18.33 & 0.03 & o & ATLAS \\
281.60 & 59597.39 & 18.41 & 0.05 & o & ATLAS \\
282.10 & 59597.90 & 18.14 & 0.22 & o & ATLAS \\
285.52 & 59601.39 & 18.73 & 0.05 & o & ATLAS \\
286.04 & 59601.92 & 20.31 & 0.03 & B & NOT+ALFOSC \\
286.04 & 59601.92 & 18.54 & 0.09 & o & ATLAS \\
286.04 & 59601.92 & 19.43 & 0.06 & V & NOT+ALFOSC \\
287.51 & 59603.42 & 18.74 & 0.10 & o & ATLAS \\
289.34 & 59605.29 & 18.86 & 0.07 & o & ATLAS \\
290.26 & 59606.22 & 18.81 & 0.05 & o & ATLAS \\
290.26 & 59606.22 & 19.60 & 0.16 & c & ATLAS \\
291.35 & 59607.33 & 18.83 & 0.04 & o & ATLAS \\
292.33 & 59608.33 & 18.67 & 0.03 & o & ATLAS \\
293.03 & 59609.05 & 19.61 & 0.05 & c & ATLAS \\
293.03 & 59609.05 & 18.87 & 0.06 & o & ATLAS \\
294.31 & 59610.35 & 18.92 & 0.04 & o & ATLAS \\
294.93 & 59610.98 & >19.75 & - & B & NOT+ALFOSC \\
294.94 & 59610.99 & 19.10 & 0.07 & i & NOT+ALFOSC \\
294.94 & 59610.99 & 18.83 & 0.05 & r & NOT+ALFOSC \\
294.94 & 59610.99 & 19.83 & 0.07 & V & NOT+ALFOSC \\
298.19 & 59614.31 & 19.89 & 0.07 & c & ATLAS \\
298.19 & 59614.31 & 18.95 & 0.15 & o & ATLAS \\
299.19 & 59615.32 & 19.09 & 0.09 & o & ATLAS \\
300.13 & 59616.29 & 19.02 & 0.05 & o & ATLAS \\
301.14 & 59617.31 & 19.16 & 0.01 & o & ATLAS \\
302.09 & 59618.29 & 18.91 & 0.08 & o & ATLAS \\
302.63 & 59618.83 & >19.75 & - & B & NOT+ALFOSC \\
302.64 & 59618.84 & 19.98 & 0.08 & V & NOT+ALFOSC \\
302.64 & 59618.84 & 19.10 & 0.05 & r & NOT+ALFOSC \\
302.64 & 59618.84 & 19.30 & 0.08 & i & NOT+ALFOSC \\
304.72 & 59620.96 & 20.08 & 0.10 & V & NOT+ALFOSC \\
304.72 & 59620.96 & 20.95 & 0.11 & B & NOT+ALFOSC \\
308.99 & 59625.32 & 18.87 & 0.10 & o & ATLAS \\
310.01 & 59626.36 & 18.93 & 0.03 & o & ATLAS \\
310.89 & 59627.25 & 18.33 & 0.18 & o & ATLAS \\
312.69 & 59629.09 & 13.20 & 0.02 & W1 & WISE \\
312.69 & 59629.09 & 19.04 & 0.06 & o & ATLAS \\
312.69 & 59629.09 & 12.66 & 0.02 & W2 & WISE \\
313.89 & 59630.31 & 19.21 & 0.03 & o & ATLAS \\
314.78 & 59631.22 & 19.41 & 0.05 & o & ATLAS \\
315.84 & 59632.29 & 19.38 & 0.07 & o & ATLAS \\

    \end{tabular}
    \label{tab:photometry_table_2}
\end{table}

\begin{table}[H]
\ContinuedFloat
\caption{continued}
    \centering
    \begin{tabular}{c c c c c c} \hline\hline
phase & mjd & mag & error & band & telescope+instrument \\ \hline
316.62 & 59633.10 & 19.48 & 0.02 & r & NOT+ALFOSC \\
316.62 & 59633.10 & 20.50 & 0.05 & V & NOT+ALFOSC \\
316.62 & 59633.10 & 21.33 & 0.07 & B & NOT+ALFOSC \\
316.62 & 59633.10 & 19.65 & 0.03 & i & NOT+ALFOSC \\
317.33 & 59633.82 & 19.76 & 0.04 & G & Gaia \\
317.41 & 59633.90 & 19.72 & 0.04 & G & Gaia \\
317.61 & 59634.11 & 19.43 & 0.26 & o & ATLAS \\
318.65 & 59635.16 & 19.37 & 0.09 & o & ATLAS \\
323.69 & 59640.30 & 19.30 & 0.24 & o & ATLAS \\
325.22 & 59641.87 & 21.64 & 0.01 & B & NOT+ALFOSC \\
325.22 & 59641.87 & 19.76 & 0.03 & r & NOT+ALFOSC \\
325.22 & 59641.87 & 19.99 & 0.05 & i & NOT+ALFOSC \\
325.22 & 59641.87 & 20.79 & 0.04 & V & NOT+ALFOSC \\
326.59 & 59643.26 & 19.08 & 0.22 & o & ATLAS \\
327.58 & 59644.27 & 19.47 & 0.10 & o & ATLAS \\
336.43 & 59653.29 & 19.05 & 0.29 & o & ATLAS \\
339.91 & 59656.84 & 20.27 & 0.12 & r & NOT+ALFOSC \\
339.91 & 59656.84 & 20.34 & 0.13 & i & NOT+ALFOSC \\
339.93 & 59656.86 & >19.75 & - & B & NOT+ALFOSC \\
339.94 & 59656.87 & >21.03 & - & V & NOT+ALFOSC \\
342.03 & 59659.00 & 18.65 & 0.23 & o & ATLAS \\
346.74 & 59663.80 & 20.54 & 0.10 & G & Gaia \\
346.81 & 59663.88 & 20.55 & 0.10 & G & Gaia \\
362.64 & 59680.01 & 20.96 & 0.02 & i & NOT+ALFOSC \\
362.64 & 59680.01 & 22.10 & 0.14 & V & NOT+ALFOSC \\
362.64 & 59680.01 & 20.70 & 0.04 & r & NOT+ALFOSC \\
365.63 & 59683.06 & 17.35 & 0.09 & J & NOT+NOTCam \\
365.63 & 59683.06 & 14.67 & 0.16 & K & NOT+NOTCam \\
365.63 & 59683.06 & 15.87 & 0.09 & H & NOT+NOTCam \\
489.47 & 59809.31 & 19.14 & 0.09 & J & NOT+NOTCam \\
489.47 & 59809.31 & 15.73 & 0.22 & K & NOT+NOTCam \\
489.47 & 59809.31 & 17.20 & 0.15 & H & NOT+NOTCam \\
499.17 & 59819.20 & 22.66 & 0.07 & r & NOT+ALFOSC \\
514.49 & 59834.82 & 13.18 & 0.04 & W2 & WISE \\
514.49 & 59834.82 & 13.95 & 0.02 & W1 & WISE \\
516.85 & 59837.22 & 19.71 & 0.12 & J & NOT+NOTCam \\
516.85 & 59837.22 & 15.98 & 0.17 & K & NOT+NOTCam \\
516.85 & 59837.22 & 17.78 & 0.10 & H & NOT+NOTCam \\
541.30 & 59862.16 & 16.28 & 0.21 & K & NOT+NOTCam \\
541.30 & 59862.16 & 18.09 & 0.09 & H & NOT+NOTCam \\
541.30 & 59862.16 & 20.50 & 0.11 & J & NOT+NOTCam \\
560.91 & 59882.15 & 18.25 & 0.10 & H & NOT+NOTCam \\
560.91 & 59882.15 & 20.32 & 0.14 & J & NOT+NOTCam \\
560.91 & 59882.15 & 16.48 & 0.18 & K & NOT+NOTCam \\
585.45 & 59907.16 & 16.60 & 0.17 & K & NOT+NOTCam \\
585.45 & 59907.16 & 20.40 & 0.11 & J & NOT+NOTCam \\
585.45 & 59907.16 & 18.32 & 0.14 & H & NOT+NOTCam \\
627.56 & 59950.10 & 18.90 & 0.12 & H & NOT+NOTCam \\
627.56 & 59950.10 & 17.07 & 0.16 & K & NOT+NOTCam \\
653.90 & 59976.95 & 18.92 & 0.10 & H & NOT+NOTCam \\
653.90 & 59976.95 & 17.31 & 0.15 & K & NOT+NOTCam \\
669.72 & 59993.08 & 13.66 & 0.03 & W2 & WISE \\
669.72 & 59993.08 & 14.55 & 0.09 & W1 & WISE \\
694.11 & 60017.94 & 19.82 & 0.21 & H & NOT+NOTCam \\
694.11 & 60017.94 & 17.78 & 0.20 & K & NOT+NOTCam \\
875.11 & 60202.47 & 14.04 & 0.01 & W2 & WISE \\
875.11 & 60202.47 & 15.62 & 0.01 & W1 & WISE \\
    \end{tabular}
    \label{tab:photometry_table_3}
\end{table}

\begin{table}[H]
\caption{Log of spectral observations of SN 2021irp}
\centering
\begin{tabular}{c c c c c c c c} \hline\hline 
Phase & Date & MJD & Telescope & Instrument & Grism & Range & Resolving power \\ 
Rest frame days & UT &  &  &  &  & \AA &    \\ 
\hline
181.5 & 2021-10-08.3 & 59495.3 & LT & SPRAT & - & 4000-8000 & 350 \\
200.0 & 2021-10-27.2 & 59514.2 & NOT & ALFOSC & Gr4 & 3200-9600 & 360  \\
203.0 & 2021-10-30.3 & 59517.3 & VLT & FORS2 & GRIS\_300V & 3800-9200 & 480 \\
245.5 & 2021-12-12.6 & 59560.6 & Seimei & KOOLS=IFT & VPH-blue & 4100-8900 & 500 \\
254.9 & 2021-12-21.2 & 59570.2 & NOT & ALFOSC & Gr4 & 3200-9600 & 360 \\
261.8 & 2021-12-29.2 & 59577.2 & VLT & FORS2 & GRIS\_1200R & 5750-7310 & 2140 \\
294.9 & 2022-01-31.9 & 59610.9 & NOT & ALFOSC & Gr7 & 3650-7110 & 500 \\
316.4 & 2022-02-22.9 & 59632.9 & NOT & ALFOSC & Gr4 & 3200-9600 & 360 \\
345.0 & 2022-03-24.0 & 59662.0 & VLT & FORS2 & GRIS\_300V & 3800-9200 & 480 \\
527.7 & 2022-09-25.3 & 59848.3 & VLT & FORS2 & GRIS\_1200R & 5750-7310 & 2140 \\
\end{tabular}
\label{tab:Spectra_log}
\end{table}

\begin{table}[H]
\caption{Host galaxy absolute magnitudes obtained using the {\sc hostphot} package  with a 6~kpc / 15.2" aperture, or from the AllWISE catalog with a 16.5" aperture in the case of WISE data. Magnitudes are corrected for the MW extinction.}
\centering
\begin{tabular}{c c c c c c c} \hline\hline 
Survey & $u$ & $g$ & $r$ & $i$ & $z$ & $y$  \\ 
 & mag & mag & mag & mag & mag & mag \\ 
\hline
PS1 & & -18.92 $\pm$ 0.02 & -19.28 $\pm$ 0.01 & -19.33 $\pm$ 0.02  & -19.49 $\pm$ 0.02 & -19.32 $\pm$ 0.02 \\
SDSS & -18.2 $\pm$ 0.3  & -19.07 $\pm$ 0.10 & -19.39 $\pm$ 0.07 & -19.53 $\pm$ 0.06  & -19.62 $\pm$ 0.06 &   \\ \hline
 & J & H & K \\
2MASS & $-20.5 \pm 0.3$ & $-21.1 \pm 0.4$ & $-21.7 \pm 0.4$ & \\ \hline
 & W1 & W2 & W3 & W4\\
WISE & $-21.67 \pm 0.03$ & $-21.63 \pm 0.03$ & $-24.71 \pm 0.06$ & $-26.53 \pm 0.5$\\
\end{tabular}
\label{tab:host_mags}
\end{table}

\begin{table}[H]
\caption{Parameters adopted and references for the comparison objects depicted in Figs \ref{fig:bolo_comp}, \ref{fig:spec_comp} and \ref{fig:Ha_spec_comp_21irp}}
\centering
\begin{tabular}{c c c c c} \hline\hline 
SN name & Redshift & Explosion date & E$(B-V)$ & Data sources  \\ \hline
SN~1998S & 0.002795 & 1998-03-02 & 0.682 & \citet{Fassia2001} \\
SN~1999em & 0.0024 & 1999-10-24 & 0.10 & \citet{Leonard2002} \\
SN~2004et & 0.0168 & 2017-10-01 & 0.03 & \citet{Maguire2010} \\
SN~2010jl & 0.01058 & 2010-10-09 & 0.058 & \citet{Jencson2016,Fransson2014,Smith2012} \\
SN~2015da & 0.0107 & 2015-01-08 & 3.04 & \citet{Tartaglia2020} \\
SN~2016gsd & 0.067 & 2016-09-17 & 0.08 & \citet{Reynolds2020} \\
SN~2017hcc & 0.0168 & 2017-10-01 & 0.030 & \citet{Moran2023} \\
SN~2017gmr & 0.00504 & 2017-09-01 & 0.30 & \citet{Andrews2019,Utrobin2021} \\
\end{tabular}
\label{tab:comparison_objects}
\end{table}

\end{appendix}
\end{document}